# On interplay of surface tension and inertial stabilization mechanisms in the interface dynamics with interfacial mass flux

D.V. Ilyin (1); S.I. Abarzhi (2)*
California Institute of Technology, Pasadena CA, 91125, USA (1);
The University of Western Australia, Perth, WA 6009, Australia (2)
dan.ilyin@gamil.com (1); snezhana.abarzhi@gmail.com (2)

This work focuses on the interfacial dynamics with interfacial mass flux in the presence of acceleration and surface tension. We employ the general matrix method to find the fundamental solutions for the linearized boundary value problem conserving mass, momentum and energy. We find that the dynamics can be stable or unstable depending on the values of the acceleration, the surface tension and the density ratio. In the stable regime, the flow has the non-perturbed fields in the bulk, is shear-free at the interface, and has the constant interface velocity. The dynamics is unstable only when it is accelerated, and when the acceleration value exceeds a threshold combining contributions of the inertial stabilization mechanism and the surface tension. The properties of this instability unambiguously differentiate it from other fluid instabilities. Particularly, its velocity field has potential and vortical components in the bulk and is shear free at the interface. Its dynamics describes the standing wave with the growing amplitude, and has the growing interface velocity. For strong accelerations, this fluid instability of the conservative dynamics has the fastest growth-rate and the largest stabilizing surface tension value when compared to the classical Landau's and Rayleigh-Taylor dynamics. We find the values of the initial perturbation wavelength at which the fluid instability can be stabilized and at which it has the fastest growth. We identify theory benchmarks for experiments and simulations in high energy density plasmas and its outcomes for application problems in nature and technology.

## Section 1 - Introduction

Non-equilibrium transport, interfaces and mixing are omnipresent in nature and technology at astrophysical and at molecular scales, and in high and low energy density regimes [1]. Fluid instabilities and interfacial mixing in supernovae and in inertial confinement fusion, particle-field interactions in magnetic fusion and in imploding Z-pinches, downdrafts in stellar interior and in planetary magneto-convection, coronal mass ejections in the Solar flares and plasma instabilities in the Earth ionosphere, plasma thrusters and nano-fabrication – are examples of processes governed by non-equilibrium interfacial dynamics [2-19]. The realistic environments are often characterized by sharply and rapidly changing flow fields and by small effects of dissipation and diffusion resulting in the formation of discontinuities (referred to as fronts or as interfaces) between the flow non-uniformities (phases) at macroscopic (continuous) scales [11]. Such processes are particularly important in high energy density plasmas, and are challenging to examine in theory, experiments and simulations [1,2,18].

In this work we systematically study the dynamics of the interface separating ideal incompressible fluids of different densities, having the interfacial mass flux, and being influenced by the acceleration and the surface tension [20]. Through the general theoretical framework [21-23], we find that the dynamics conserving mass, momentum and energy is stabilized by the inertial mechanism and by surface tension and is destabilized by acceleration. For large accelerations the dynamics is unstable, leading to the growth of the interface perturbations and the growth of the interface velocity. This instability of the conservative dynamics can be unambiguously discerned from other fluid instabilities.

This paper is organized as follows. We start with the Introduction in Section 1. We provide the Method in Section 2, including the governing equations (2.1), the theoretical approach (2.2), and the fundamental solutions (2.3). We present in Section 3 the Results of our analysis for the conservative dynamics, the classical Landau's dynamics, and Rayleigh-Taylor dynamics influenced by the acceleration and the surface tension. This includes the fundamental solutions (3.1), the systematic study of the properties of the inertial dynamics free from surface tension (3.2) and with surface tension (3.3), the focused analysis of the accelerated dynamics free from surface tension (3.4) and with surface tension (3.5), the investigations of the mechanisms of the interface stabilization and destabilization (3.6) and the characteristic scales (3.7), as well as theory outcomes for experiments and simulations (3.8). We finalize the work with Discussion in Section 4, and provide Acknowledgements, Data availability, Author's contributions, References, Tables, and Figure captions and Figures in Sections 5-10.

When looking from a far field, an observer ordinarily considers two kinds of discontinuities separating the flow phases: a front and an interface [20]. The front has zero mass flux across it. Through the interface the mass can be transported. The fluid phases are broadly defined: These can be the distinct kinds of matter or the same kind of matter with distinct properties. To describe the multi-phase flow, a boundary value problem is solved by balancing the fluxes of mass, momentum and energy at the freely evolving discontinuity. While boundary value problems are challenging to investigate, this approach has a number of advantages, and its solution has high predictive capability in a broad parameter regime [21].

The unstable accelerated fronts are represented by Rayleigh-Taylor and Richtmyer-Meshkov instabilities [24-27]. The fundamental properties of Rayleigh-Taylor / Richtmyer-Meshkov dynamics in the scale-dependent early-time and late-time regimes and in the self-similar interfacial mixing regime are well captured by the group theory and by the linear and weakly nonlinear theories [28-31]. For interfaces, the classical theoretical framework for the problem was developed by Landau [32]. It considered the dynamics of ideal incompressible fluids, balanced at the interface the fluxes of mass and momentum and postulated the special condition for the perturbed mass flux [32]. Several seminal models further connected this framework to realistic environments in high energy density plasmas and in reactive and super-critical fluids [33-35,36-38].

The dynamics of interfaces with interfacial mass flux is a long-standing problem in science, mathematics and engineering [1]. It has wide-ranging applications in plasmas (dynamics of ablation front influencing the hot spot formation in inertial confinement fusion), astrophysics (thermonuclear flashes determining the nuclear synthesis in type-Ia supernova), material science (material transformations under high strain rates in nano-fabrication), and industry (scramjets) [1,2]. To tackle these research frontiers and solve a broad class of problems, the theory of interface dynamics was recently developed [21-23].

This theory elaborated the general framework for the problem of interface stability, directly linked the microscopic transport at the interface to macroscopic fields in the fluids' bulk, and reported the mechanisms of the interface stabilization and destabilization never previously discussed [21-23]. The key discoveries – the inertial mechanism of the interface stabilization, the new fluid instability of the accelerated interface, and the chemistry-induced instability – identified the fundamental properties of the interface dynamics. They also resolved the long-standing prospect of Landau [32], by showing that the classical Landau's solution for the Landau-Darrieus instability is a perfect mathematical match [21-23].

The theory [21-23] considered inertial and accelerated interface dynamics for ideal incompressible fluids free from stabilizations caused by interactions of particles at molecular scales [14-17]. Realistic processes are usually accompanied by dissipation, diffusion, compressibility, radiation transport, stratification, and surface tension [3-15,39-43]. The influence of these effects on the interface dynamics call for systematic investigations [21].

Here we study the interfacial dynamics with interfacial mass flux in the presence of acceleration and surface tension. The fluids are ideal and incompressible, with negligible stratification and densities variation, the flow is two-dimensional, periodic and spatially extended. The acceleration is directed from the heavy to the light fluid. The surface tension is understood as a tension at the phase boundary between the flow phases [20,21]. At microscopic scales the surface tension results from the imbalance of forces of intermolecular interactions near the interface [14,20,41]. Physically, it is always present in a multiphase flow. Mathematically, it is accounted for through the pressure modification in the governing equations [14,20,21]. We investigate the interplay of macroscopic and microscopic stabilization mechanisms (due to the inertial effect and surface tension, respectively) and the destabilizing effect of the acceleration.

In order to address this task, we advance and employ the general matrix method to rigorously solve the linearized boundary value problem conserving mass, momentum and energy [21,22]. We find that depending on the acceleration and surface tension, the dynamics can be stable or unstable. In the stable regime, the flow has non-perturbed flow fields in the bulk, is shear-free at the interface and has the constant interface velocity. The conservative dynamics with the surface tension is unstable only when it is accelerated and when the acceleration value exceeds a threshold. The threshold value combines the contributions of the inertial and the surface tension mechanisms and is finite for zero surface tension. In the unstable regime, the dynamics couples the interface perturbations with the potential and vortical components of the velocity fields in the fluids' bulk and is shear free at the interface. It describes the standing wave with the growing amplitude, and has the growing interface velocity. This instability of the conservative dynamics has unique quantitative and qualitative properties unambiguously differentiating it from other fluid instabilities. Particularly, it has the fastest growth-rate and the largest stabilizing surface tension value in the extreme regime of strong accelerations, when compared to the accelerated Landau's and Rayleigh-Taylor dynamics with surface tension. We further find the critical and maximum values of the initial perturbation wavelengths at which the fluid instability of the conservative dynamics can be stabilized and at which its growth is the fastest. Based on the obtained results, we identify the theory benchmarks for future experiments and simulations in high energy density plasmas and outline its outcomes for application problems in nature and technology [1-15,39-45].

## Section 2 – Method

### Sub-Section 2.1 - Governing equations

In the inertial frame of reference, the dynamics of ideal fluid is governed by the conservation of mass, momentum, and energy as

$$\frac{\partial \rho}{\partial t}+\frac{\partial \rho v_i}{\partial x_i}=0, \quad \frac{\partial \rho v_i}{\partial t}+\frac{\partial \rho v_i v_j}{\partial x_j}+\frac{\partial P}{\partial x_i}=0, \quad \frac{\partial E}{\partial t}+\frac{\partial (E+P)v_i}{\partial x_i}=0 \quad (1)$$

Here $x_i$ are the spatial coordinates, $(x_1, x_2, x_3) = (x, y, z)$, $t$ is time, $(\rho, \mathbf{v}, P, E)$ are the fields of density $\rho$, velocity $\mathbf{v}$, pressure $P$ and energy $E = \rho(e + \mathbf{v}^2/2)$, and $e$ is specific internal energy [20-23]. The inertial frame of reference is referred to the frame of reference moving with constant velocity $\tilde{\mathbf{V}}_0$; for definiteness $\tilde{\mathbf{V}}_0 = (0, 0, \tilde{V}_0)$ [21,22].

For a system of two fluids with different densities separated by an interface, we mark the fields of the heavy (light) fluid as $(\rho, \mathbf{v}, P, E)_{h(l)}$, and we introduce a continuous local scalar function $\theta(x, y, z, t)$ to describe the interface. The function value is $\theta = 0$ at the interface and it is $\theta > 0$ ($\theta < 0$) in the heavy (light) fluid [21-23,28,29]. By using the Heaviside step-function $H(\theta)$ we represent the flow fields in the entire domain as $(\rho, \mathbf{v}, P, E) = (\rho, \mathbf{v}, P, E)_h H(\theta) + (\rho, \mathbf{v}, P, E)_l H(-\theta)$ [19,21-23,18,19].

At the interface, the balance of fluxes of mass and normal and tangential components of momentum and energy obey the boundary conditions [21-23,28,29]:

$$[\tilde{\mathbf{j}} \cdot \mathbf{n}] = 0, \quad \left[\left(P + \frac{(\tilde{\mathbf{j}} \cdot \mathbf{n})^2}{\rho}\right)\mathbf{n}\right] = 0, \quad \left[\frac{(\tilde{\mathbf{j}} \cdot \mathbf{n})(\tilde{\mathbf{j}} \cdot \boldsymbol{\tau})}{\rho}\boldsymbol{\tau}\right] = 0, \quad \left[(\tilde{\mathbf{j}} \cdot \mathbf{n})\left(W + \frac{\tilde{\mathbf{j}}^2}{2\rho^2}\right)\right] = 0 \quad (2)$$

where the jump of functions across the interface is denoted with $[...]$; the unit vectors normal and tangential at the interface are $\mathbf{n}$ and $\boldsymbol{\tau}$ with $\mathbf{n} = \nabla\theta/|\nabla\theta|$ and $(\mathbf{n} \cdot \boldsymbol{\tau}) = 0$; the mass flux across the moving interface is $\tilde{\mathbf{j}} = \rho(\mathbf{n}\dot{\theta}/|\nabla\theta| + \mathbf{v})$; the specific enthalpy is $W = e + P/\rho$ [21-23,28,29].

We consider the spatially extended flow, which is unbounded in the $z$ direction and is periodic in the $(x, y)$ plane. The heavy (light) fluid is located in the lower (upper) sub-domain. The boundary conditions at the outside boundaries of the domain are

$$\mathbf{v}_h\big|_{z \to -\infty} = \mathbf{V}_h = (0, 0, V_h), \quad \mathbf{v}_l\big|_{z \to +\infty} = \mathbf{V}_l = (0, 0, V_l) \quad (3)$$

with the constant velocity magnitude(s) $V_{h(l)}$, Figure 1.

The interface velocity in the inertial frame of reference is $\tilde{\mathbf{V}}$. For the steady planar interface normal to the mass flux, the interface velocity is constant; this velocity can be chosen equal to the velocity of the inertial frame of reference as $\tilde{\mathbf{V}}_0 = \tilde{\mathbf{V}}$. For the non-steady non-planar interface arbitrarily positioned relative to the mass flux, the interface velocity $\tilde{\mathbf{V}}$ and the velocity of the inertial frame of reference $\tilde{\mathbf{V}}_0$ are distinct, $\tilde{\mathbf{V}} \neq \tilde{\mathbf{V}}_0$ [21]. In this general case, the interface velocity $\tilde{\mathbf{V}}$ obeys the relation

$$\tilde{\mathbf{V}}\mathbf{n} = -\mathbf{v}\mathbf{n}\big|_{\theta \to 0^+} = -(\tilde{\mathbf{j}}/\rho)\mathbf{n}\big|_{\theta \to 0^+} \qquad (4)$$

The flow is subject to the acceleration and the interfacial surface tension. The acceleration $\mathbf{g}$ is directed along the $z$ direction from the heavy fluids to the light fluid, as $\mathbf{g} = (0,0,g)$. The interfacial surface tension is understood as the tension between the fluid phases, and is characterized by the surface tension coefficient $\sigma, \sigma \geq 0$.

We consider a sample case of a two-dimensional flow periodic in the $x$ direction, free from motion in the $y$ direction and spatially extended in the $z$ direction. The interfacial function $\theta$ is set as

$$\theta = -z + z^*(x,t), \quad \dot{\theta} = \frac{\partial z^*}{\partial t}, \quad \nabla\theta = \left(\frac{\partial z^*}{\partial x}, 0, -1\right), \quad |\nabla\theta| = \sqrt{1 + \left(\frac{\partial z^*}{\partial x}\right)^2} \quad (5)$$

**Sub-Section 2.2 - Linearized dynamics**

The unperturbed flow fields are $\{\tilde{\mathbf{j}}, \mathbf{v}, P, W\} = \{\mathbf{J}, \mathbf{V}, P_0, W_0\}$, and the normal and tangential unit vectors of the unperturbed interface are $\{\mathbf{n}, \boldsymbol{\tau}\} = \{\mathbf{n}_0, \boldsymbol{\tau}_0\}$ Eqs.(2) We slightly perturb in Eqs.(1-4) the flow fields as $\tilde{\mathbf{j}} = \mathbf{J} + \mathbf{j}$, $\mathbf{v} = \mathbf{V} + \mathbf{u}$, $P = P_0 + p$, and $W = W_0 + w$, with $|\mathbf{j}| << |\mathbf{J}|$, $|\mathbf{u}| << |\mathbf{V}|$ $|p| << |P_0|$ and $|w| << |W_0|$. We slightly perturb the fluid interface as $\mathbf{n} = \mathbf{n}_0 + \mathbf{n}_1$ and $\boldsymbol{\tau} = \boldsymbol{\tau}_0 + \boldsymbol{\tau}_1$, with $|\mathbf{n}_1| << |\mathbf{n}_0|$ and $|\boldsymbol{\tau}_1| << |\boldsymbol{\tau}_0|$, and with $|\dot{\theta}/|\nabla\theta|| << |\mathbf{V}|$. The fluid density is perturbed as $\rho \to \rho + \delta\rho$ with $|\delta\rho| << |\rho|$. The perturbed velocity of the interface is $\tilde{\mathbf{V}} = \tilde{\mathbf{V}}_0 + \tilde{\mathbf{v}}$, with $|\tilde{\mathbf{v}}| << |\tilde{\mathbf{V}}_0|$.

To the leading order in small perturbations, the boundary conditions at the interface are:

$$[\mathbf{J}\cdot\mathbf{n}_0] = 0, \quad \left[\left(P_0 + \frac{(\mathbf{J}\cdot\mathbf{n}_0)^2}{\rho}\right)\mathbf{n}_0\right] = 0, \quad \left[(\mathbf{J}\cdot\mathbf{n}_0)\frac{(\mathbf{J}\cdot\boldsymbol{\tau}_0)}{\rho}\boldsymbol{\tau}_0\right] = 0, \quad \left[(\mathbf{J}\cdot\mathbf{n}_0)\left(W_0 + \frac{\mathbf{J}^2}{2\rho^2}\right)\right] = 0 \quad (6.1)$$

To the first order, the boundary conditions at the interface are:

$$[\mathbf{j}\cdot\mathbf{n}_0 + \mathbf{J}\cdot\mathbf{n}_1] = 0, \quad \left[\left(p + \frac{2(\mathbf{J}\cdot\mathbf{n}_0)}{\rho}\left((\mathbf{j}\cdot\mathbf{n}_0 + \mathbf{J}\cdot\mathbf{n}_1) - \frac{(\mathbf{J}\cdot\mathbf{n}_0)}{2}\frac{\delta\rho}{\rho}\right)\right)\mathbf{n}_0\right] = 0, \quad (6.2)$$

$$\left[((\mathbf{J}\cdot\mathbf{n}_0)(\mathbf{J}\cdot\boldsymbol{\tau}_1 + \mathbf{j}\cdot\boldsymbol{\tau}_0) + (\mathbf{J}\cdot\boldsymbol{\tau}_0)(\mathbf{J}\cdot\mathbf{n}_1 + \mathbf{j}\cdot\mathbf{n}_0))\frac{\boldsymbol{\tau}_0}{\rho}\right] = 0,$$

$$\left[(\mathbf{J}\cdot\mathbf{n}_0)\left(w + \frac{(\mathbf{J}\cdot\mathbf{j})}{\rho^2} - \left(\frac{\delta\rho}{\rho}\right)\frac{\mathbf{J}^2}{\rho^2}\right)\right] = 0$$

The small perturbations of the flow fields decay away from the interface:

$$\{\mathbf{j},\mathbf{v},p,w,\delta\rho\}_h\big|_{z\to-\infty}=0,\quad \{\mathbf{j},\mathbf{v},p,w,\delta\rho\}_l\big|_{z\to+\infty}=0 \quad (6.3)$$

The interface velocity is $\tilde{\mathbf{V}}=\tilde{\mathbf{V}}_0+\tilde{\mathbf{v}}$, and, up to the first order, it is

$$\tilde{\mathbf{V}}=\tilde{\mathbf{V}}_0+\tilde{\mathbf{v}},\quad \tilde{\mathbf{v}}\mathbf{n}_0=-\left(\mathbf{u}_h\mathbf{n}_0+\dot{\theta}\right)\Big|_{\theta=0^+} \qquad (6.4)$$

The boundary conditions in Eqs.(2-4) and Eqs.(6) are valid for compressible and incompressible ideal fluids, for two- and three-dimensional flows, and for arbitrary positioning of the interface relative the mass flux. The conditions Eqs.(6) can be simplified by applying the conditions of directionality of the mass flux, the incompressibility of the fluids, and the dimensionality of the flow.

Indeed, to the leading order, the mass flux is $\mathbf{J}=\rho\mathbf{V}$, the flow fields are uniform in the bulk, $(\rho,\mathbf{v},P,W)_{h(l)}=(\rho,\mathbf{V},P_0,W_0)_{h(l)}$, and obey conditions Eqs.(3) at the boundaries of the domain. The components of mass flux normal and tangential to the interface are $J_n=\mathbf{J}\cdot\mathbf{n}_0$, $J_\tau=\mathbf{J}\cdot\boldsymbol{\tau}_0$. We presume that the leading order mass flux is normal to the planar interface; hence, its tangential component is zero, $J_\tau=0$. In the limiting case of incompressible dynamics, the values approach $\left(P_0+J_n^2/\rho\right)_{h(l)}\to\left(P_0\right)_{h(l)}$ and $\left(W_0+\mathbf{J}^2/2\rho^2\right)_{h(l)}\to\left(W_0\right)_{h(l)}$, since the speed(s) of sound in the fluid(s) is substantially greater than other velocity scales. These transform Eqs.(6.1) to

$$[J_n]=0,\ [P_0\,\mathbf{n}_0]=0,\ [J_n\,W_0]=0 \quad (7)$$

For a two-dimensional flow in Eqs.(5), to the leading order the normal and tangential vectors of the interface are $\mathbf{n}_0=(0,0,-1)$ and $\boldsymbol{\tau}_0=(1,0,0)$. The first order perturbations of the normal and tangential vectors of the interface are $\mathbf{n}_1=\left(\partial z^*/\partial x,0,0\right)$ and $\boldsymbol{\tau}_1=\left(0,0,\partial z^*/\partial x\right)$. This leads to $\mathbf{J}\cdot\mathbf{n}_1=0$. For incompressible fluids with negligible density perturbations $|\delta\rho/\rho|<<|\mathbf{u}|/|\mathbf{V}|$ the first order boundary conditions at the interfaces Eqs.(6.2) are then transformed to

$$[\mathbf{j}\cdot\mathbf{n}_0]=0,\quad \left[\left(\left(p+\frac{2(\mathbf{J}\cdot\mathbf{n}_0)}{\rho}\right)(\mathbf{j}\cdot\mathbf{n}_0)\right)\mathbf{n}_0\right]=0,\quad (8.1)$$

$$\left[(\mathbf{J}\cdot\mathbf{n}_0)\frac{(\mathbf{J}\cdot\boldsymbol{\tau}_1+\mathbf{j}\cdot\boldsymbol{\tau}_0)}{\rho}\boldsymbol{\tau}_0\right]=0,\quad \left[(\mathbf{J}\cdot\mathbf{n}_0)\left(w+\frac{(\mathbf{J}\cdot\mathbf{j})}{\rho^2}\right)\right]=0$$

The normal and tangential components of the perturbed mass flux are $j_n=\mathbf{j}\cdot\mathbf{n}_0$ and $j_\tau=\mathbf{j}\cdot\boldsymbol{\tau}_0$. The perturbed flow fields in the bulk and at the outside boundaries are [20-23]:

$$\nabla\cdot\mathbf{u}_{h(l)}=0,\quad \dot{\mathbf{u}}_{h(l)}+\left(\mathbf{V}_{h(l)}\cdot\nabla\right)\mathbf{u}_{h(l)}+\frac{\nabla p_{h(l)}}{\rho_{h(l)}}=0,\quad \mathbf{u}_h\big|_{z\to-\infty}=0,\quad \mathbf{u}_l\big|_{z\to+\infty}=0 \quad (8.2)$$

**Section 2.3 - Fundamental solutions**

We seek solutions for the boundary value problem Eqs.(8) in which the perturbed velocity of the heavy fluid is potential in accordance with the Kelvin theorem, and the perturbed velocity of the light fluid has both potential and vortical components [20-22,23]:

$$\mathbf{u}_h = \nabla\Phi_h, \quad \mathbf{u}_l = \nabla\Phi_l + \nabla\times\mathbf{\Psi}_l \qquad (9.1)$$

The fluid potential and vortical fields and the interface perturbation are

$$\Phi_h = \Phi\, exp(ikx + kz + \Omega t), \quad \Phi_l = \tilde{\Phi}\, exp(ikx - kz + \Omega t) \qquad (9.2)$$
$$\mathbf{\Psi}_l = (0, \Psi_l, 0), \quad \Psi_l = \Psi\, exp\left(ikx - \tilde{k}z + \Omega t\right), \quad z^* = Z\, exp(ikx + \Omega t)$$

Here $\Omega$ is the growth-rate (the characteristic frequency, the eigenvalue) of the system equations Eqs.(8), $k = 2\pi/\lambda$ is the wavevector and $\lambda$ is the spatial period (the wavelength).

For the pressure perturbations $p_{h(l)}$ and for the length-scale of the vortical field $\tilde{\lambda} = 2\pi/\tilde{k}$, we obtain from Eqs.(8.2)

$$\nabla\left(\dot{\Phi}_{h(l)} + V_{h(l)}\left(\frac{\partial\Phi_{h(l)}}{\partial z}\right) + \frac{p_{h(l)}}{\rho_{h(l)}}\right) = 0, \quad \left(\frac{\partial}{\partial t} + V_l\frac{\partial}{\partial z}\right)(\nabla\times\mathbf{\Psi}_l) = 0 \qquad (9.3)$$

The perturbed pressure is free from contributions from the perturbed vortical field [20,21]. This leads to

$$p_{h(l)} = -\rho_{h(l)}\left(\dot{\Phi}_{h(l)} + V_{h(l)}\frac{\partial\Phi_{h(l)}}{\partial z}\right), \qquad \tilde{k} = \Omega/V_l \qquad (9.4)$$

In order to obey the boundary conditions $\mathbf{u}_l\big|_{z\to+\infty} = 0$, the vortical field should decay away from the interface, $\left(\tilde{k}/k\right) > 0$, and the interface dynamics should be unstable, $\Omega(kV_h) > 0$.

In the presence of acceleration $\mathbf{g} = (0, 0, g)$, the pressure perturbations are modified as [20-23]:

$$p_{h(l)} = -\rho_{h(l)}\left(\dot{\Phi}_{h(l)} + V_{h(l)}\frac{\partial\Phi_{h(l)}}{\partial z} - gz\right) \qquad (9.5.1)$$

In the presence of surface tension the pressure perturbations are further modified as [20,21]:

$$(p_h - p_l) \to (p_h - p_l) + \sigma\frac{\partial^2 z^*}{\partial x^2} \qquad (9.5.2)$$

With expressions Eqs.(9), the system of differential equations governing the interface dynamics is reduced to the linear system $\mathrm{M}\mathbf{r} = 0$, where vector $\mathbf{r}$ is $\mathbf{r} = \left(\Phi_h, \Phi_l, V_h z^*, \Psi_l\right)^{\mathrm{T}}$, and the matrix $\mathrm{M}$ is defined by the boundary conditions at the interface Eqs.(8,9) [21].

For ideal incompressible fluids, the characteristic length-scale $1/k$ and time-scale $1/kV_h$ are defined by the initial conditions, and the characteristic density scale is set by the heavy fluid density $\rho_h$. We use the dimensionless values of the growth-rate $\omega = \Omega/kV_h$, and the density ratio $R = \rho_h/\rho_l$ with $R \geq 1$. This leads to $V_l/V_h = R$, $\tilde{k}/k = \omega/R$ [21-23]. We use the dimensional values of the gravity $G = g/kV_h^2$, with $G \geq 0$, and surface tension $T = (\sigma/\rho_h)(k/V_h^2)$, with $T \geq 0$ [20-22]. We use dimensionless values for the flow fields, the interface, and the variables as $\varphi = \Phi/(V_h/k)$, $\tilde{\varphi} = \tilde{\Phi}/(V_h/k)$, $\psi = \Psi/(V_h/k)$, $\bar{z} = kZ$, and $kx \to x, kz \to z, kV_h t \to t$. In the dimensionless units, the fluid potentials are $\varphi_h = \varphi e^{ix+z+\omega t}$, $\varphi_l = \tilde{\varphi} e^{ix-z+\omega t}$ and $\boldsymbol{\psi}_l = (0, \psi_l, 0)$ with $\psi_l = \psi e^{ix-(\tilde{k}/k)z+\omega t}$, the fluid velocities are $\mathbf{u}_h = \nabla\varphi_h$, $\mathbf{u}_l = \nabla\varphi_l + \nabla\times\boldsymbol{\psi}_l$, and the interface perturbation is $z^* = \bar{z} e^{ix+\omega t}$.

In the dimensionless form, the elements of the matrix $\mathrm{M}$ is are the functions of the growth-rate (the frequency, the eigenvalue) $\omega$, the density ratio $R$, the acceleration value $G$, and the surface tension value $T$ as $\mathrm{M} = \mathrm{M}(\omega, R, G, T)$ [21]. The condition $det\,\mathrm{M}(\omega_i, R, G, T) = 0$ defines the eigenvalues $\omega_i$. The associated eigenvectors $\tilde{\mathbf{e}}_i$ are derived by reducing the matrix $\mathrm{M}(\omega_i, R, G, T)$ to row-echelon form [21,22]. The matrix $\mathrm{M}$ is $4\times 4$. For a non-degenerate $4\times 4$ matrix, there are 4 fundamental solutions $\mathbf{r}_i = \mathbf{r}_i(\omega_i, \tilde{\mathbf{e}}_i)$, $i = 1...4$, with 4 associated eigenvalues $\omega_i$ and eigenvectors $\tilde{\mathbf{e}}_i$, corresponding to 4 degrees of freedom and 4 independent variables obeying 4 equations, Eqs.(8,9).

Solution $\mathbf{r}$ for the system $\mathrm{M}\mathbf{r} = 0$ is a linear combination of the fundamental solutions $\mathbf{r}_i$

$$\mathbf{r} = \sum_{i=1}^{4} C_i \mathbf{r}_i \qquad (10)$$

Here $C_i$ are the integration constants, and $\mathbf{r}_i = \mathbf{r}_i(\omega_i, \tilde{\mathbf{e}}_i)$ are the fundamental solutions with $\mathbf{r}_i = \tilde{\mathbf{e}}_i e^{\omega_i t}$ and $\tilde{\mathbf{e}}_i = \left(\varphi e^{ix+z}, \tilde{\varphi} e^{ix-z}, \bar{z} e^{ix}, \psi e^{ix-(\tilde{k}/k)z}\right)_i^{\mathrm{T}}$ and with the associated vector $\mathbf{e}_i = (\varphi, \tilde{\varphi}, \bar{z}, \psi)_i^{\mathrm{T}}$.

By using the general matrix method for solving the boundary value problem Eqs.(1-10), we directly link the microscopic interfacial transport to macroscopic flow fields, and we conduct a systematic study of the interplay of the interface stability with the structure of the flow fields [21,22].

## Section 3 - Results

### Sub-Section 3.1 - Matrixes and fundamental solutions

In this section we identify the fundamental solutions for the accelerated conservative dynamics and for the classical Landau's and Rayleigh-Taylor dynamics with the acceleration and surface tension.

### Sub-Section 3.1.1 - Conservative dynamics

We consider the conservative dynamics balancing the fluxes of mass, momentum and energy at the interface, Eqs.( 8.1). For this dynamics, the matrix $\mathrm{M}$ is $\mathrm{M}=\mathrm{M}_{GT}$:

$$\mathrm{M}_{GT} = \begin{pmatrix} -R & -1 & -\omega + R\omega & i \\ 1 & -1 & 1-R & i\omega/R \\ R - R\omega & R+\omega & G(R-1)-RT & -2iR \\ \omega & -\omega & T+\omega-R\omega & iR \end{pmatrix} \quad (11.1)$$

Its determinant is $det\mathrm{M}_{GT} = i((R-1)/R)(\omega-R)(\omega+R)(\omega^2(R-1)+R(R-1)-G(R+1)+TR)$, and the values $\omega_i$ and $\mathbf{e}_i$ are

$$\omega_{1(2)} = \pm i\sqrt{R}\sqrt{1-\frac{G(R+1)}{R(R-1)}+\frac{T}{R-1}}, \quad \mathbf{e}_{1(2)} = (\varphi, \tilde{\varphi}, 1, \psi)^{\mathrm{T}}_{1(2)}; \qquad (11.2)$$

$$\omega_3 = R, \quad \mathbf{e}_3 = (0, i, 0, 1)^{\mathrm{T}}; \quad \omega_4 = -R, \quad \mathbf{e}_4 = \left(\frac{2i}{R+1}, -\frac{i(R-1)}{R+1}, 0, 1\right)^{\mathrm{T}}$$

where the components $\{\varphi, \tilde{\varphi}, \psi\}$ of the eigenvectors for solutions 1 and 2 are functions on $R, G, T$. Among the fundamental solutions for the conservative dynamics Eqs.(11), the fundamental solutions $\mathbf{r}_1(\omega_1, \mathbf{e}_1)$ and $\mathbf{r}_2(\omega_2, \mathbf{e}_2)$ depend on the values of the acceleration $G$ and the surface tension $T$, whereas the fundamental solutions $\mathbf{r}_3(\omega_3, \mathbf{e}_3)$ and $\mathbf{r}_4(\omega_4, \mathbf{e}_4)$ are independent of $G$ and $T$ [21,22].

In regards to the fundamental solutions $\mathbf{r}_1(\omega_1, \mathbf{e}_1)$ and $\mathbf{r}_2(\omega_2, \mathbf{e}_2)$ in Eqs.(11), for some values of the acceleration, the surface tension and the density ratio, these solutions are stable, with $\mathbf{r}_1 = \mathbf{r}_2^*$ and $\omega_1 = \omega_2^*$ with $\mathrm{Re}[\omega_{1(2)}] = 0$. They describe two stable traveling waves, whose superposition results in stably oscillating standing waves. For some other values of the acceleration, the surface tension and the density ratio, one of these solutions is unstable, $\mathbf{r}_1$ with $\mathrm{Re}[\omega_1] > 0$, whereas the other is stable, $\mathbf{r}_2$ with $\mathrm{Re}[\omega_2] > 0$. These solutions describe the standing waves, with the growing ($\mathbf{r}_1$) and the decaying ($\mathbf{r}_2$) amplitudes. For these solutions, the interface perturbations are coupled with the potential and vortical components of the velocities of the fluids' bulk.

In regards to the fundamental solutions $\mathbf{r}_3(\omega_3, \mathbf{e}_3)$ and $\mathbf{r}_4(\omega_4, \mathbf{e}_4)$ in Eqs.(11), the solution $\mathbf{r}_3$ is unstable, $\omega_3 = R$ and $\mathrm{Re}[\omega_3] > 0$, and the solution $\mathbf{r}_4$ is stable, $\omega_4 = -R$ and $\mathrm{Re}[\omega_4] < 0$. The remarkable property of the formally unstable solution $\mathbf{r}_3$ is that the interface perturbation and the perturbed fields of the velocities and pressure are identically zero in the entire domain at any time for any integration constant $C_3$, with $z^* = 0$, $\mathbf{u}_{h(l)} = 0$, $p_{h(l)} = 0$ [21,22]. For the formally stable fundamental solution $\mathbf{r}_4$, we must set the integration constant $C_4 = 0$, in order for this solution to obey at any time the conditions $\mathbf{u}_l\big|_{z\to+\infty} = 0$. This is because the vortical component of the velocity, $\nabla\times\boldsymbol{\psi}_l \neq 0$, while decaying in time, increases away from the interface. Note that for solution $\mathbf{r}_4$ the vorticity value is $\nabla\times\mathbf{u}_l = 0$, despite of $\boldsymbol{\psi}_l \neq 0$ and $\nabla\times\boldsymbol{\psi}_l \neq 0$. This is because in the vorticity field $\nabla\times\mathbf{u}_l = \left(0, \left(1-\left(\tilde{k}/k\right)^2\right)\psi_l, 0\right)$ the values are $\left(\tilde{k}/k\right)^2 = (\omega/R)^2 = 1$ [21,22].

The accelerated conservative dynamics with surface tension has 4 fundamental solutions with 4 associated eigenvalues and eigenvectors, 4 independent degrees of freedom, and is non-degenerate. By defining the solution $\mathbf{r}_{CDGT}$ in the stable regime as the superposition of the traveling waves $\mathbf{r}_{CDGT} = (\mathbf{r}_1 + \mathbf{r}_2)/2$ and in the unstable regime as the solution $\mathbf{r}_{CDGT} = \mathbf{r}_1$, we analyze properties of this solution below, Table 1. Sub-script stands for conservative dynamics with the gravity $G$ and the surface tension $T$.

**Sub-Section 3.1.2 – Classical Landau's dynamics**

The classical Landau's dynamics balances the fluxes of mass, and normal and tangential components of momentum, and employs the special condition for the perturbed velocity at the interface. This special condition postulates the constancy of the interface velocity of the non-steady non-planar interface, $\tilde{\mathbf{V}} \equiv \tilde{\mathbf{V}}_0$, and leads to [20-23,32-35]:

$$[\mathbf{j}\cdot\mathbf{n}_0] = 0, \quad \left[\left(\left(p + \frac{2(\mathbf{J}\cdot\mathbf{n}_0)}{\rho}\right)(\mathbf{j}\cdot\mathbf{n}_0)\right)\mathbf{n}_0\right] = 0, \quad (12.1)$$

$$\left[(\mathbf{J}\cdot\mathbf{n}_0)\frac{(\mathbf{J}\cdot\boldsymbol{\tau}_1 + \mathbf{j}\cdot\boldsymbol{\tau}_0)}{\rho}\boldsymbol{\tau}_0\right] = 0, \quad [\mathbf{u}\cdot\mathbf{n}_0] = 0$$

For the Landau's dynamics the matrix $\mathrm{M}$ is $\mathrm{M} = \mathrm{L}_{GT}$.

$$\mathrm{L}_{GT} = \begin{pmatrix} -R & -1 & -\omega + R\omega & i \\ 1 & -1 & 1-R & i\omega/R \\ R - R\omega & R+\omega & G(R-1) - RT & -2iR \\ -1 & -1 & 0 & i \end{pmatrix} \quad (12.2)$$

Its determinant is $det\,\mathrm{L}_{GT} = i((R-1)/R)(\omega - R)((R+1)\omega^2 + 2R\omega - (R-1)(R+G) + TR)$, and the values of $\omega_i$ and $\mathbf{e}_i$ are:

$$\omega_{1(2)} = \frac{-R \pm \sqrt{(R^3 + R^2 - R) + G(R^2 - 1) - TR(R+1)}}{R+1}, \quad \mathbf{e}_{1(2)} = (\varphi, \tilde{\varphi}, 1, \psi)^{\mathrm{T}}_{1(2)}; \quad (12.3)$$

$$\omega_3 = R, \mathbf{e}_3 = (0, i, 0, 1)^{\mathrm{T}}$$

where the components of eigenvectors $\{\varphi, \tilde{\varphi}, \psi\}$ for solutions 1 and 2 are functions on $R, G, T$ .

Among the fundamental solutions for the classical Landau's dynamics, the fundamental solutions $\mathbf{r}_1(\omega_1, \mathbf{e}_1)$ and $\mathbf{r}_2(\omega_2, \mathbf{e}_2)$ depend on the values of the acceleration $G$ and surface tension $T$, and the fundamental solution $\mathbf{r}_3(\omega_3, \mathbf{e}_3)$ is independent of $G$ and $T$ and is identical to that in Eqs.(11) [21,22].

For the classical Landau's dynamics Eqs.(12), the fundamental solution $\mathbf{r}_1(\omega_1, \mathbf{e}_1)$ corresponds to Landau-Darrieus instability in the gravity field in the presence of the surface tension. For this solution, the interface perturbations are coupled the potential and vortical components of the velocities in the fluids' bulk. For the fundamental solution $\mathbf{r}_2(\omega_2, \mathbf{e}_2)$ the interface perturbation and the potential and vortical components of the velocities are also coupled. For this solution we must set the integration constant $C_2 = 0$, in order to obey at any time the condition $\mathbf{u}_l|_{z \to +\infty} = 0$ in Eqs.(8). Solution $\mathbf{r}_3(\omega_3, \mathbf{e}_3)$ has zero fields of the perturbed velocity and pressure in the entire domain for any integration constant $C_3$ and at any time, as in Eqs.(12) [21,22].

The accelerated Landau's dynamics with the surface tension is degenerate, since it has smaller number of fundamental solutions (3) than the number of the degrees of freedom (4). This indicates a singular and ill-posed character of the dynamics. The lifting the degeneracy may lead to a scale-invariant power-law dynamics and be triggered by a seed vortical field, pre-imposed in the bulk of the light fluid at some instance of time [21].

By defining the solution as $\mathbf{r}_{LDGT} = \mathbf{r}_1$, we analyze properties of this solution below, Table 2. Sub-script stands for Landau's dynamics with the gravity and the tension.

**Sub-Section 3.1.3 - Rayleigh-Taylor dynamics**

The boundary conditions Eqs.(2-4) are derived from the governing equations Eqs.(1) assuming that the mass flux is conserved at the interface, $\left[\tilde{\mathbf{j}}\cdot\mathbf{n}\right]=0$. There is the important particular case, when the conserved mass flux is zero at the interface, $\tilde{\mathbf{j}}\cdot\mathbf{n}\big|_{\theta=0}=0$. This leads to the continuity of normal component of velocity $[\mathbf{v}\cdot\mathbf{n}]=0$, the continuity of the pressure, and the arbitrariness of the jumps of tangential component of velocity and enthalpy at the interface [19,20-22,28,29]:

$$[\mathbf{v}\cdot\mathbf{n}]=0, \quad [P]=0, \quad [\mathbf{v}\cdot\boldsymbol{\tau}]=arbitrary, \quad [W]=arbitrary \qquad (13.1)$$

For the zero mass flux at the interface, the outside boundaries have no influence on the dynamics:

$$\mathbf{v}\big|_{z\to+\infty}=0, \quad \mathbf{v}\big|_{z\to-\infty}=0 \quad (13.2)$$

and the interface velocity is zero in the laboratory frame of reference:

$$\tilde{\mathbf{V}}=0 \qquad (13.3)$$

This case corresponds to the dynamics of contact discontinuity and Rayleigh-Taylor instability.

For Rayleigh-Taylor dynamics, the unperturbed interface is planar, and the unperturbed velocity field is zero in both fluids. We slightly perturb the interface as $\theta=-z+z^*(x,t)$, with $z^*=Ze^{ikx+\Omega t}$, $\left|\dot{\theta}/|\nabla\theta|\right|<<\sqrt{g/k}$ and $\left|\partial z^*/\partial x\right|<<1$. We slightly perturb the fluid velocities with the potential fields, $\mathbf{v}_h=\nabla\Phi_h$, $\Phi_h=\Phi e^{ikx+kz+\Omega t}$, and $\mathbf{v}_l=\nabla\Phi_l$, $\Phi_l=\tilde{\Phi}e^{ikx-kz+\Omega t}$, with $|\mathbf{v}|<<\sqrt{g/k}$. We perturb the fluid pressure as $P=P_0+p$, $|p|<<|P_0|$, with $p_{h(l)}=-\rho_{h(l)}\left(\dot{\Phi}_{h(l)}+V_{h(l)}\,\partial\Phi_{h(l)}/\partial z-gz\right)$ and further modify it as $(p_h-p_l)\to(p_h-p_l)+\sigma\left(\partial^2z^*/\partial x^2\right)$. System Eqs.(13) is then reduced to a linear system $\mathrm{M}\,\mathbf{r}=0$, where vector $\mathbf{r}$ is $\mathbf{r}=\left(\Phi_h,\Phi_l,V_hz^*\right)^{\mathrm{T}}$ and $\mathrm{M}$ is the $3\times3$ matrix.

In Rayleigh-Taylor dynamics the length-scale is $1/k$ and the time-scale is $1/\sqrt{gk}$. In order to conduct a comparative study of this dynamics with the conservative dynamics and the classical Landau's dynamics, we scale the time with $1/kV_h$, where $V_h$ is now understood as some velocity scale. This leads to $G=g/kV_h^2$ and $T=(\sigma/\rho_h)\left(k/V_h^2\right)$, as before. In the dimensional units the matrix $\mathrm{M}=\mathrm{M}(\omega,R,G,T)$. For system $\mathrm{M}\,\mathbf{r}=0$, the solution is $\mathbf{r}=\sum_i C_i\mathbf{r}_i$, with $i=1,2,3$ in non-degenerate case, similarly to Eqs.(10). Here $C_i$ are the integration constants, $\mathbf{r}_i=\mathbf{r}_i(\omega_i,\tilde{\mathbf{e}}_i)$ are the fundamental solutions with $\mathbf{r}_i=\tilde{\mathbf{e}}_ie^{\omega_it}$, $\tilde{\mathbf{e}}_i=\left(\varphi e^{ix+z},\tilde{\varphi}e^{ix-z},\bar{z}e^{ix}\right)_i^{\mathrm{T}}$ are the eigenvectors, and $\mathbf{e}_i = (\varphi_i, \tilde{\varphi}_i, \bar{z}_i)^{\mathrm{T}}$ are the associated vectors. For Rayleigh-Taylor dynamics in Eqs.(13), matrix $\mathrm{M}$ is $\mathrm{M} = \mathrm{T}_{GT}$:

$$\mathrm{T}_{GT} = \begin{pmatrix} -R & -1 & -\omega + R\omega \\ -R - R\omega & -R + \omega & G(R-1) - TR \\ -1 & -1 & 0 \end{pmatrix} \quad (14.1)$$

Its determinant is $det\,\mathrm{T}_{GT} = (R-1)\left((R+1)\omega^2 - G(R-1) + TR\right)$, and $\omega_i$ and $\mathbf{e}_i$ are

$$\omega_{1(2)} = \pm\sqrt{\frac{G(R-1) - TR}{R+1}}, \quad \mathbf{e}_{1(2)} = (\varphi, \tilde{\varphi}, 1)^{\mathrm{T}}_{1(2)} \quad (14.2)$$

where the components of eigenvectors $\{\varphi, \tilde{\varphi}\}$ are the functions on $R, G, T$ .

Depending on the values of the acceleration, the surface tension and the density ratio, the solutions $\mathbf{r}_1(\omega_1, \mathbf{e}_1)$ and $\mathbf{r}_2(\omega_2, \mathbf{e}_2)$ can be stable or unstable. When both solutions are stable, with $\mathbf{r}_1 = \mathbf{r}_2^*$ and $\omega_1 = \omega_2^*$ with $\mathrm{Re}[\omega_{1(2)}] = 0$, they describe traveling waves whose superposition results in stably oscillating standing waves. For some other values of the acceleration, the surface tension and the density ratio, one of these solutions is unstable, $\mathbf{r}_1$ with $\mathrm{Re}[\omega_1] > 0$, whereas the other is stable, $\mathbf{r}_2$ with $\mathrm{Re}[\omega_2] > 0$. These solutions describe the standing waves, with the growing ($\mathbf{r}_1$) and decaying ($\mathbf{r}_2$) amplitudes. For solutions $\mathbf{r}_1$ and $\mathbf{r}_2$ the velocity fields are potential in the fluids' bulk. For $G > 0, R > 1, T = 0$, solution $\mathbf{r}_1(\omega_1, \mathbf{e}_1)$ corresponds to Rayleigh-Taylor instability [19-21,24-31].

Rayleigh-Taylor dynamics is degenerate, with smaller number of fundamental solutions (2) than the degrees of freedom (3), and, hence, it is singular and ill-posed. The lifting the degeneracy may lead to a power-law dynamics. Such dynamics can be triggered by a seed vortical field pre-imposed at the interface at some instance of time (e.g., vortex line or the vortex sheet). This happens in, e.g., Richtmyer-Meshkov instability, due to the vorticity deposition at the interface and the impulsive acceleration by the shock [21,22,24-31].

By defining the solution as $\mathbf{r}_{RTGT} = (\mathbf{r}_1 + \mathbf{r}_2)/2$ in the stable regime, and as $\mathbf{r}_{RTGT} = \mathbf{r}_1$ in the unstable regime, we analyze properties of this solution below, Table 3. Sub-script stands for Rayleigh-Taylor dynamics with the gravity and tension.

**Sub-Section 3.2 – Inertial dynamics free from surface tension**

In this sub-section, for the purpose of completeness, we provide solutions $\{\mathbf{r}_{CDGT}, \mathbf{r}_{LDGT}, \mathbf{r}_{RTGT}\}$ for inertial dynamics free from surface tension, $G = 0, T = 0$, see Ref.[21] for details.

For the conservative dynamics the solution is $\mathbf{r}_{CDGT}\,|_{G=0,T=0} = \mathbf{r}_{CDGT}\left(\omega_{CDGT}, \tilde{\mathbf{e}}_{CDGT}\right)|_{G=0,T=0}$:

$$\omega_{CDGT}\big|_{G=0,T=0} = \pm i\sqrt{R}, \quad \mathbf{e}_{CDGT}\big|_{G=0,T=0} = \frac{\mathbf{e}+\mathbf{e}^*}{2}, \quad \mathbf{e} = \left(\varphi, \tilde{\varphi}, 1, 0\right)^{\mathrm{T}} \quad (15.1)$$
$$\tilde{\mathbf{V}} = \tilde{\mathbf{V}}_0 + \tilde{\mathbf{v}}, \quad \tilde{\mathbf{v}}\mathbf{n}_0 = -\left(\mathbf{u}_h\mathbf{n}_0 + \dot{\theta}\right)\big|_{\theta=0} \sim e^{\pm i\sqrt{R}t}$$

The components of the eigenvector are $\varphi = i(R-1)/\left(i+\sqrt{R}\right)$ and $\tilde{\varphi} = -(R-1)\sqrt{R}/\left(i+\sqrt{R}\right)$. This solution is stable. It is stabilized by the inertial mechanism. Mathematically, the mechanism is revealed in stable oscillations of the interface velocity near the constant value $\tilde{\mathbf{V}} = \tilde{\mathbf{V}}_0 + \tilde{\mathbf{v}}$, with $\tilde{\mathbf{v}}\cdot\mathbf{n}_0 \sim e^{\pm i\sqrt{R}t}$. Physically, when the interface is perturbed, the parcels of the heavy fluid and the light fluid follow the interface perturbation thus causing the change of momentum and energy of the fluid system. Yet, the dynamics is inertial. To conserve the momentum and energy, the interface as whole should slightly change its velocity. This causes the reactive force to occur and stabilize the dynamics [21,22].

For classical Landau's dynamics, the solution is $\mathbf{r}_{LDGT}\,|_{G=0,T=0} = \mathbf{r}_{LDGT}\left(\omega_{LDGT}, \mathbf{e}_{LDGT}\right)|_{G=0,T=0}$

$$\omega_{LDGT}\big|_{G=0,T=0} = \frac{-R+\sqrt{-R+R^2+R^3}}{1+R}, \quad \mathbf{e}_{LDGT}\big|_{G=0,T=0} = \mathbf{e}, \quad \mathbf{e} = \left(\varphi, \tilde{\varphi}, 1, \psi\right)^{\mathrm{T}} \quad (15.2)$$
$$\tilde{\mathbf{V}} \equiv \tilde{\mathbf{V}}_0, \quad \tilde{\mathbf{V}} = \tilde{\mathbf{V}}_0 + \tilde{\mathbf{v}}, \quad \tilde{\mathbf{v}}\mathbf{n}_0 = -\left(\mathbf{u}_h\mathbf{n}_0 + \dot{\theta}\right)\big|_{\theta=0} \equiv 0$$

The components of the eigenvector $\{\varphi, \tilde{\varphi}, \psi\}$ are the functions on the density ratio $R$. This solution is unstable. When the interface is perturbed, the parcels of the heavy fluid and the light fluid follow the interface perturbation thus causing the change of momentum and energy of the fluid system. Yet, the postulated constancy of the interface velocity, $\tilde{\mathbf{V}} \equiv \tilde{\mathbf{V}}_0$, which is implemented in the special boundary condition $[\mathbf{u}\mathbf{n}_0] \equiv 0$, preempts the occurrence of the reactive force. The interface perturbations grow and Landau-Darrieus instability develops [21,32].

For Rayleigh-Taylor dynamics the solution is $\mathbf{r}_{RTGT}\,|_{G=0,T=0} = \mathbf{r}_{RTGT}\left(\omega_{RTGT}, \tilde{\mathbf{e}}_{RTGT}\right)|_{G=0,T=0}$

$$\omega_{RTGT}\big|_{G=0,T=0} = 0, \quad \mathbf{e}_{CDGT}\big|_{G=0,T=0} = \mathbf{e}, \quad \mathbf{e} = \left(0, 0, 1\right)^{\mathrm{T}}, \quad \tilde{\mathbf{V}} = 0 \quad (15.3)$$

This solution is neutrally stable. It has zero interface velocity in the laboratory frame of reference [21].

Hence, for the inertial dynamics free from surface tension $\left\{\mathbf{r}_{CDGT}, \mathbf{r}_{LDGT}, \mathbf{r}_{RTGT}\right\}_{G=0,T=0}$: The conservative dynamics is stable; it has potential flow fields in the fluids' bulk and is shear free at the interface; it is stabilized by the inertial mechanism revealed in stable oscillations of the interface velocity near the constant value. The classical Landau's dynamics is unstable; it has potential and vortical components of the velocity in the fluids' bulk; it is shear free at the interface; it has the postulated constant interface velocity; Rayleigh-Taylor dynamics is neutrally stable; it has zero velocity fields in the fluids' bulk; it has zero interface velocity in the laboratory frame of reference.

**Sub-Section 3.3 – Inertial dynamics with surface tension**

Here we investigate solutions $\{\mathbf{r}_{CDGT}, \mathbf{r}_{LDGT}, \mathbf{r}_{RTGT}\}$ in the case of the inertial dynamics with the surface tension, $G = 0, T > 0$. The results are illustrated by Figures 2,3 and Tables 1-4.

**Sub-Section 3.3.1 – Conservative dynamics**

For the conservative dynamics the solution is $\mathbf{r}_{CDGT}\big|_{G=0} = \mathbf{r}_{CDGT}\left(\omega_{CDGT}, \tilde{\mathbf{e}}_{CDGT}\right)\big|_{G=0}$ with

$$\omega_{CDGT}\big|_{G=0} = \pm i\sqrt{R}\sqrt{1+\frac{T}{R-1}}, \quad \mathbf{e}_{CDGT}\big|_{G=0} = \frac{\mathbf{e}+\mathbf{e}^{*}}{2}, \quad \mathbf{e} = \left(\varphi, \tilde{\varphi}, 1, \psi\right)^{\mathrm{T}} \quad (16.1)$$

where quantities $\{\varphi, \tilde{\varphi}, \psi\}$ are the functions on the density ratio and surface tension $R, T$. This solution is consistent with the solution for the inertial conservative dynamics free from surface tension Eqs.(15.1), Table 1, with $(\varphi, \tilde{\varphi}, 1, \psi) \to (\varphi, \tilde{\varphi}, 1, 0)$ for $T \to 0$.

The flow field for solution $\mathbf{r}_{CDGT}\big|_{G=0}$ have the following structure Eqs.(16.1). For the inertial conservative dynamics with finite surface tension value, the velocity field is potential in the heavy fluid bulk, and has potential and vortical components in the light fluid bulk Eqs.(16.1), Figure 3, Table 1. The appearance of the vortical field in the light fluid bulk is associated with the contribution of surface energy, which defines the strength of the vortical field. In the limit of zero surface tension, the velocity fields are potential in both fluids.

The inertial dynamics with the surface tension $\mathbf{r}_{CDGT}\big|_{G=0}$ is stable for $R > 1, T > \tilde{T}_{cr}\big|_{G=0}$, $\tilde{T}_{cr}\big|_{G=0} = 0$, Figure 2, Table 4. The eigenvalue $\omega_{CDGT}\big|_{G=0}$ is imagine, $\mathrm{Re}\left[\omega_{CDGT}\big|_{G=0}\right] = 0$. This suggests that the length-scale of the vortical field $\tilde{k} = (k/R)\,\omega_{CDGT}\big|_{G=0}$ is also imagine, $\mathrm{Re}\left[\tilde{k}\right] = 0$. Hence, the dynamics $\mathbf{r}_{CDGT}\big|_{G=0}$ describes the standing wave stably oscillating in time, Figure 2. For this wave, in the bulk of the heavy fluid the velocity field is potential; it decays away from the interface. In the bulk of the light fluid, the velocity field has potential and vortical components. Its potential component decays away from the interface. The vortical field is periodic in the $x$ direction with the period $\lambda = 2\pi/k$, and is also periodic in the $z$ direction with the period $\tilde{\lambda} = 2\pi/\tilde{k}$. Hence, this dynamics has the stably oscillating periodic vortical structure with constant amplitude, Figure 3. For solution $\mathbf{r}_{CDGT}\big|_{G=0}$ the vorticity $\nabla\times\mathbf{u}_l=\left(0,\left(1-\left(\tilde{k}/k\right)^2\right)\psi_l,0\right)$ is $\nabla\times\mathbf{u}_l\neq 0$; its field is also periodic in the $(x,z)$ plane, Figure 3. We see that in order to obey the boundary condition $\mathbf{u}_l\big|_{z\to+\infty}=0$ in Eqs.(8.2) for the solution $\mathbf{r}_{CDGT}\big|_{G=0}$ in Eqs.(11,16.1) we must set its integration constant equal zero $C_{CDGT}\big|_{G=0}=0$.

Consider now the interplay of the surface tension with the inertial stabilization mechanism. Since the solution is periodic in time, one might expect that the interface velocity experiences stable oscillations, similarly to the case of the inertial dynamics free from surface tension [21]. However, since the integration constant is zero, the interface velocity for this solution is constant:

$$\tilde{\mathbf{V}}=\tilde{\mathbf{V}}_0+\tilde{\mathbf{v}},\quad \tilde{\mathbf{v}}\mathbf{n}_0=-\left(\mathbf{u}_h\mathbf{n}_0+\dot{\theta}\right)\Big|_{\theta=0}\sim e^{\pm i\left|\omega_{CDGT}\right|_{G=0}\left|t\right.},\quad C_{CDGT}\big|_{G=0}=0\quad\Rightarrow\quad\tilde{\mathbf{V}}=\tilde{\mathbf{V}}_0\quad(16.2)$$

We see that for the conservative dynamics, the inertial stabilization mechanism is present, since the inertial dynamics is stable. This mechanism is however 'masked' by the surface tension for $T>\tilde{T}_{cr}\big|_{G=0}$, since the integration constant for the solution is zero. It is exhibited only at $T=\tilde{T}_{cr}\big|_{G=0}=0$.

Therefore, the inertial conservative dynamics with the surface tension is stable for any values of the density ratio and the surface tension $R>1, T>\tilde{T}_{cr}\big|_{G=0}$, $\tilde{T}_{cr}\big|_{G=0}=0$, Figure 2, Tables 1,4. For the solution $\mathbf{r}_{CDGT}\big|_{G=0}$ the integration constant must be zero $C_{CDGT}\big|_{G=0}=0$ in order to satisfy the boundary conditions away from the interface, Figure 3. The resultant inertial conservative dynamics of the interface with surface tension corresponds to the stable unperturbed flow fields $(\rho,\mathbf{V},P_0,W_0)_{h(l)}$ and has constant interface velocity $\tilde{\mathbf{V}}=\tilde{\mathbf{V}}_0$.

**Sub-Section 3.3.2 – Classical Landau's dynamics**

For the classical Landau's dynamics the solution is $\mathbf{r}_{LDGT}\big|_{G=0}=\mathbf{r}_{LDGT}\left(\omega_{LDGT},\tilde{\mathbf{e}}_{LDGT}\right)\big|_{G=0}$ with

$$\omega_{LDGT}\big|_{G=0}=\frac{-R+\sqrt{\left(R^3+R^2-R\right)-TR(R+1)}}{R+1},\quad \mathbf{e}_{LDGT}\big|_{G=0}=\mathbf{e}=\left(\varphi,\tilde{\varphi},1,\psi\right)^{\mathrm{T}}\quad(17)$$

where the quantities $\{\varphi,\tilde{\varphi},\psi\}$ are the functions on the density ratio and the surface tension $R,T$, Table 2. This solution is consistent with the solution for classical Landau's dynamics free from surface tension, since for $T\to 0$ components $\{\varphi,\tilde{\varphi},\psi\}_{LDGT|_{G=0}}\to\{\varphi,\tilde{\varphi},\psi\}_{LDGT|_{G=0,T=0}}$ in agreement with Eqs.(15.2).

The solution $\mathbf{r}_{LDGT}\big|_{G=0}$ is stable for $T > \overline{T}_{cr}\big|_{G=0}$, and is unstable for $T < \overline{T}_{cr}\big|_{G=0}$, where $\overline{T}_{cr}\big|_{G=0} = R - 1$, in agreement with Ref.[20,21], Figure 2, Table 4.

The investigation of properties of the solution $\mathbf{r}_{LDGT}\big|_{G=0}$ for $T > \overline{T}_{cr}\big|_{G=0}$ suggests that in the stable regime, its integration constant must be set zero $C_{LDGT}\big|_{G=0} = 0$ in order to obey the boundary conditions far from the interface.

Consider properties of the solution $\mathbf{r}_{LDGT}\big|_{G=0}$ in the unstable regime, for $T < \overline{T}_{cr}\big|_{G=0}$, Figure 2, Table 4. This solution corresponds to Landau-Darrieus instability with surface tension, and satisfies the assigned boundary conditions at the interface and at the outside boundaries of the domain Eqs.(12). Its dynamics couples the interface perturbation with the vortical and potential components of the velocity fields. For solution $\mathbf{r}_{LDGT}\big|_{G=0}$ the vortical component of the velocity of the light fluid $\nabla \times \boldsymbol{\psi}_l$ and the vorticity $\nabla \times \mathbf{u}_l$, while increasing in time, decay far from the interface. The vortical field has the wavevector $\widetilde{k} = (k/R)\,\omega_{LDGT}\big|_{G=0}$ and the length-scale $\widetilde{\lambda}/\lambda = k/\widetilde{k}$, Figure 2, Table 2. The interface velocity for the solution $\mathbf{r}_{LDGT}\big|_{G=0}$ is constant, $\widetilde{\mathbf{V}} \equiv \widetilde{\mathbf{V}}_0$, in both stable and unstable regimes, as postulated by the interfacial boundary conditions in the classical Landau's dynamics, Eqs.(12).

**Sub-Section 3.3.3 – Rayleigh-Taylor dynamics**

For the inertial Rayleigh-Taylor dynamics of contact discontinuity with surface tension the solution is $\mathbf{r}_{RTGT}\big|_{G=0} = \mathbf{r}_{RTGT}\left(\omega_{RTGT}, \widetilde{\mathbf{e}}_{RTGT}\right)\big|_{G=0}$:

$$\omega_{RTGT}\big|_{G=0} = \pm i\sqrt{\frac{T\,R}{R+1}}, \quad \mathbf{e}_{CDGT}\big|_{G=0} = \frac{\mathbf{e}+\mathbf{e}^*}{2}, \quad \mathbf{e} = (\varphi, \widetilde{\varphi}, 1)^{\mathrm{T}} \quad (18)$$

This solution is stable. For $T > \hat{T}_{cr}\big|_{G=0}$, $\hat{T}_{cr}\big|_{G=0} = 0$, the solution corresponds to a standing capillary wave stably oscillating in time. At $T = 0$ the solution is neutrally stable, and the components of this solution are $\{\varphi, \widetilde{\varphi}\}_{RTGT|_{G=0}} = \{0, 0\}_{RTGT|_{G=0,\,T=0}}$, Figure 2, Tables 3,4.

**Sub-Section 3.3.4 – Summary of properties**

Compare the properties of the solutions $\{\mathbf{r}_{CDGT}, \mathbf{r}_{LDGT}, \mathbf{r}_{RTGT}\}_{G=0}$ for inertial dynamics with surface tension, Figure 2, Figure 3, Tables 1-4.

The conservative dynamics is stable for surface tension values $T > \tilde{T}_{cr}\big|_{G=0}, \overline{T}_{cr}\big|_{G=0} = 0$. The presence of the surface tension ‘masks’ the inertial stabilization mechanism. The resultant dynamics $\mathbf{r}_{CDGT}\ |_{G=0}$ corresponds to the stable unperturbed flow fields, and has constant interface velocity.

The classical Landau’s dynamics is stable for $T > \overline{T}_{cr}\big|_{G=0}$ and is unstable for $T < \hat{T}_{cr}\big|_{G=0}$ with $\overline{T}_{cr}\big|_{G=0} = R-1$. In the stable regime, the dynamics corresponds to the unperturbed flow fields. In the unstable regime, it couples the interface perturbation to the potential and vortical components of the velocity fields in the fbulk and it is shear-free at the interface. The classical Landau’s dynamics postulates the constancy of the interface velocity.

The inertial Rayleigh-Taylor dynamics is stable for $T > \hat{T}_{cr}\big|_{G=0}, \hat{T}_{cr}\big|_{G=0} = 0$. In the stable regime it describes the stably oscillating capillary wave. The dynamics has potential velocity fields in the bulk and has the interfacial shear. The interface velocity is zero in the laboratory frame of reference.

**Sub-Section 3.4 – Accelerated dynamics free from surface tension**

In this sub-section, for the purpose of completeness, we briefly provide the properties of solutions $\{\mathbf{r}_{CDGT}, \mathbf{r}_{LDGT}, \mathbf{r}_{RTGT}\}$ for the accelerated dynamics free from surface tension, $G > 0, T = 0$ [21,22]. The details can be found in [21,22].

For the conservative dynamics the solution is $\mathbf{r}_{CDGT}\ |_{T=0} = \mathbf{r}_{CDGT}\left(\omega_{CDGT}, \tilde{\mathbf{e}}_{CDGT}\right)\big|_{T=0}$

$$G < G_{cr}, \quad \omega_{CDGT}\big|_{T=0} = \pm i\sqrt{R}\sqrt{1-\frac{G}{G_{cr}}}, \quad \mathbf{e}_{CDGT}\big|_{T=0} = \frac{\mathbf{e}+\mathbf{e}^*}{2}, \quad \mathbf{e} = \left(\varphi, \tilde{\varphi}, 1, 0\right)^{\mathrm{T}}; \quad (19.1)$$

$$G > G_{cr}, \quad \omega_{CDGT}\big|_{T=0} = \sqrt{R}\sqrt{\frac{G}{G_{cr}}-1}, \quad \mathbf{e}_{CDGT}\big|_{T=0} = \mathbf{e} = \left(\varphi, \tilde{\varphi}, 1, 0\right)^{\mathrm{T}}, \quad G_{cr} = \frac{R(R-1)}{R+1};$$

$$\tilde{\mathbf{V}} = \tilde{\mathbf{V}}_0 + \tilde{\mathbf{v}}, \quad \tilde{\mathbf{v}}\mathbf{n}_0 = -\left(\mathbf{u}_h\mathbf{n}_0 + \dot{\theta}\right)\big|_{\theta=0} \sim e^{\left(\omega_{CDGT}|_{T=0}\right)t}$$

where quantities $\{\varphi, \tilde{\varphi}\}$ are the functions on the density ratio and the acceleration $R, G$, and $G_{cr}$ is the critical threshold value of the acceleration. For $G \to 0$, this solution is consistent with the solution for the inertial dynamics free from surface tension Eqs.(15.1). For $G > 0$ the solution’s stability is defined by the interplay of the buoyancy and the inertia, or the gravity and the reactive force. For small acceleration values, $G < G_{cr}$, the inertial effect dominates, and the reactive force exceeds the gravity. The solution is stable, and describes the standing wave stably oscillating in time. The flow dynamics is similar to the case of the inertial conservative dynamics free from surface tension. For large acceleration values, $G > G_{cr}$, the buoyant effect dominates, and the gravity exceeds the reactive force. The solution is unstable, and describes the standing wave with the growing amplitude. For this solution the velocity field is potential in the bulk, and is shear free at the interface. The flow is the superposition of two motions – the motion of the interface as whole with the growing interface velocity and the growth of the interface perturbations [21,22].

For the accelerated Landau's dynamics the solution is $\mathbf{r}_{LDGT}\big|_{T=0} = \mathbf{r}_{LDGT}\left(\omega_{LDGT}, \tilde{\mathbf{e}}_{LDGT}\right)\big|_{T=0}$

$$\omega_{LDGT}\big|_{T=0} = \frac{-R \pm \sqrt{\left(R^3 + R^2 - R\right) + G\left(R^2 - 1\right)}}{R+1}, \quad \mathbf{e}_{LDGT}\big|_{T=0} = \mathbf{e} = \left(\varphi, \tilde{\varphi}, 1, \psi\right)^{\mathrm{T}} \quad (19.2)$$

$$\tilde{\mathbf{V}} = \tilde{\mathbf{V}}_0 + \tilde{\mathbf{v}}, \quad \tilde{\mathbf{v}}\mathbf{n}_0 = -\left(\mathbf{u}_h \mathbf{n}_0 + \dot{\theta}\right)\Big|_{\theta=0} \equiv 0, \quad \tilde{\mathbf{V}} \equiv \tilde{\mathbf{V}}_0$$

where quantities $\{\varphi, \tilde{\varphi}, \psi\}$ are the functions on the density ratio and the acceleration $R, G$. For $G \to 0$, this solution is consistent with that for the inertial Landau's dynamics free from surface tension Eqs.(15.2). This solution is unstable for $G \geq 0$ and describes the standing wave with the growing amplitude. Its velocity field is potential in the bulk of the heavy fluid, has vortical and potential components in the bulk of the light fluid, and is shear free at the interface. The flow is the superposition of two motions – the motion of the interface with the postulated constant velocity and the growth of the interface perturbations [20-22].

For Rayleigh-Taylor dynamics the solution is $\mathbf{r}_{RTGT}\big|_{T=0} = \mathbf{r}_{RTGT}\left(\omega_{RTGT}, \tilde{\mathbf{e}}_{RTGT}\right)\big|_{T=0}$

$$\omega_{RTGT}\big|_{T=0} = \sqrt{\frac{G(R-1)}{R+1}}, \quad \mathbf{e}_{RTGT}\big|_{T=0} = \mathbf{e} = \left(\varphi, \tilde{\varphi}, 1\right)^{\mathrm{T}}, \quad \tilde{\mathbf{V}} \equiv 0 \quad (19.3)$$

where quantities $\{\varphi, \tilde{\varphi}\}$ are the functions on the density ratio and the acceleration $R, G$. This solution is unstable for any $G > 0$ and describes the standing wave with the growing amplitude. Its velocity field is potential in the bulks of the heavy and the light fluids, and has shear at the interface. In the laboratory frame of reference the interface velocity is zero [20-22].

A brief comparison of properties of the solutions $\{\mathbf{r}_{CDGT}, \mathbf{r}_{LDGT}, \mathbf{r}_{RTGT}\}_{T=0}$ in Eqs.(19) suggests: The accelerated conservative dynamics is unstable when the acceleration magnitude exceeds a threshold value set by inertial stabilization mechanism, $G > G_{cr}$; it has the growing interface velocity in the unstable regime; it has potential flow fields in the fluids' bulk, and is shear free at the interface. The accelerated Landau's dynamic is unstable for the acceleration values $G \geq 0$; it has a postulated constant interface velocity preempting the inertial stabilization mechanism to occur; it has a potential velocity field in the heavy fluid bulk and potential and vortical velocity fields in the light fluid bulk; it is shear free at the interface. Rayleigh-Taylor dynamics is unstable for any acceleration value $G > 0$ ; it has zero interface velocity in the laboratory frame of reference; it has potential velocity fields in the fluids' bulk, and it has the interfacial shear. For large acceleration values $G > G^*, G^* = (R^2 - 1)/4$ the instability of the accelerated conservative dynamics has the largest growth-rate when compared to the cases of the Landau-Darrieus and Rayleigh-Taylor instabilities, see for details [20-22,23-30].

**Sub-Section 3.5 – Accelerated dynamics with surface tension**

In this section we investigate the properties of the solutions $\{\mathbf{r}_{CDGT}, \mathbf{r}_{LDGT}, \mathbf{r}_{RTGT}\}$ for the accelerated dynamics with surface tension, $G > 0, T > 0$, Figures 4-9, Tables 1-3,5,6.

**Sub-Section 3.5.1 – Conservative dynamics**

**Fundamental solution**: For the accelerated conservative dynamics with the surface tension the solution is $\mathbf{r}_{CDGT} = \mathbf{r}_{CDGT}(\omega_{CDGT}, \tilde{\mathbf{e}}_{CDGT})$, Figures 4-7, Tables 1,5:

$$\tilde{G} < G_{cr}, \quad \omega_{CDGT} = \pm i\sqrt{R}\sqrt{1 - \frac{\tilde{G}}{G_{cr}}}, \quad \mathbf{e}_{CDGT} = \frac{\mathbf{e} + \mathbf{e}^*}{2}, \quad \mathbf{e} = (\varphi, \tilde{\varphi}, 1, \psi)^{\mathrm{T}}; \quad (20)$$

$$\tilde{G} > G_{cr}, \quad \omega_{CDGT} = \sqrt{R}\sqrt{\frac{\tilde{G}}{G_{cr}} - 1}, \quad \mathbf{e}_{CDGT} = \mathbf{e} = (\varphi, \tilde{\varphi}, 1, \psi)^{\mathrm{T}};$$

$$\tilde{G} = G - \frac{T\,R}{R+1}, \quad G_{cr} = \frac{R(R-1)}{R+1};$$

$$\tilde{\mathbf{V}} = \tilde{\mathbf{V}}_0 + \tilde{\mathbf{v}}, \quad \tilde{\mathbf{v}}\mathbf{n}_0 = -(\mathbf{u}_h\mathbf{n}_0 + \dot{\theta})\Big|_{\theta=0} \sim e^{\left|\omega_{CDGT}|_{T=0}\right| t}$$

where the modified acceleration $\tilde{G}$ accounts the contribution of the surface tension $T$, and $G_{cr}$ is the critical threshold acceleration value in the zero surface tension case. The quantities $\{\varphi, \tilde{\varphi}, \psi\}$ depend on the density ratio, the acceleration and the surface tension $R, G, T$. The vortical field is $\psi = -iT\,R(R-1)/(G(R+1) - R(R^2 + T - 1))$, with $\psi|_{T=0} = 0$. This solution is consistent with the solution Eqs.(19.1) for the accelerated conservative dynamics free from surface tension.

**Stability and instability of the fundamental solution:** For $G > 0$ the stability of the solution $\mathbf{r}_{CDGT}$ is defined by the interplay of the buoyancy, the inertia and the surface tension, or - the gravity, the reactive force and the tension force. The stability curve is defined by the condition $\tilde{G} = G_{cr}$ balancing the buoyancy (the gravity) with the combined contributions of the inertial stabilization mechanism and the surface tension (the reactive force and the tension force). For small acceleration values, $0 < \tilde{G} < G_{cr}$, the buoyancy is dominated and the reactive and tension forces exceed the gravity. The solution is stable, and describes the standing wave stably oscillating in time. The flow dynamics is similar to the case of the inertial conservative dynamics with surface tension. For large acceleration values, $\tilde{G} > G_{cr}$, the buoyant effect dominates and the gravity exceeds the reactive and the tension forces. The solution is unstable, and describes the standing wave with the growing amplitude, Figures 4,5, Tables 1,5.

For given values of the density ratio and the surface tension, the solution is stable for $0 < G < \tilde{G}_{cr}$ and is unstable for $G > \tilde{G}_{cr}$, where the threshold value is $\tilde{G}_{cr} = R(R-1+T)/(R+1)$, with $\tilde{G}_{cr} \to G_{cr}$ for $T \to 0$. For given values of the density ratio and the acceleration, the solution is stable for $T > \tilde{T}_{cr}$, and is unstable for $T < \tilde{T}_{cr}$, where the critical surface tension value is $\tilde{T}_{cr} = (G(R+1) - R(R-1))/R$; it approaches $\tilde{T}_{cr} \to 0$ for $G \to G_{cr}^{+}$ and equals zero $\tilde{T}_{cr} = 0$ for $0 \le G \le G_{cr}$, Figures 4,5,6,7, Tables 1,5.

**Structure of the flow fields:** Consider the structure of the flow fields for the solution $\mathbf{r}_{CDGT}$ Eqs.(20), Table 1, Figure 6,7. In this solution, in the limit of zero surface tension, $T \to 0$, the vortical component vanishes, $\psi \to 0$, and the accelerated conservative dynamics free from surface tension has potential velocity fields in the fluids' bulks [21,22]. For a finite value of the surface tension, $T > 0$, in the solution $\mathbf{r}_{CDGT}$, the vortical component is finite, $\psi \neq 0$, Eqs.(20), Table 1, Figure 6,7. This accelerated conservative dynamics with surface tension has potential velocity field in the bulk of the heavy fluid, and the velocity field combining the potential and vortical components in the bulk of the light fluid. The appearance of the vortical field in the light fluid bulk is due to the surface energy contribution to the enthalpy jump at the interface $[w]$, and it defines the strength of the vortical field.

**Stable dynamics:** For $G < \tilde{G}_{cr}$ $(T > \tilde{T}_{cr})$ the accelerated conservative dynamics with surface tension $\mathbf{r}_{CDGT}$ is stable, Eqs.(20), Figure 4-6, Tables 1,5. In this regime, the eigenvalue $\omega_{CDGT}$ is purely imagine, $\mathrm{Re}[\omega_{CDGT}] = 0$. The length-scale of the vortical field $\tilde{k} = (k/R)\,\omega_{CDGT}$ is also purely imagine, $\mathrm{Re}\,[\tilde{k}\,] = 0$. The solution $\mathbf{r}_{CDGT}$ is the standing wave stably oscillating in time, Figure 6. In the heavy fluid bulk, the velocity field is potential; it is periodic in the $x$ direction with the period $\lambda = 2\pi/k$ and decays away from the interface $z \to -\infty$. In the bulk of the light fluid, the velocity field combines the potential and the vortical components. The potential component is periodic in the $x$ direction with period $\lambda = 2\pi/k$ and decays away from the interface $z \to +\infty$. The vortical component is periodic in the $x$ direction with period $\lambda = 2\pi/k$, and is periodic in the $z$ direction with period $\tilde{\lambda} = 2\pi/\tilde{k}$. The vorticity is $\nabla \times \mathbf{u}_l = \left(0, \left(1 - \left(\tilde{k}/k\right)^2\right)\psi_l, 0\right)$ and $\nabla \times \mathbf{u}_l \neq 0$, and the amplitude of this vortical structure is constant, Figure 6. In order to obey for the solution $\mathbf{r}_{CDGT}$ in the stable regime, $0 < G < \tilde{G}_{cr}$ the boundary condition $\mathbf{u}_l\big|_{z \to +\infty} = 0$, we must set its integration constant equal zero, $C_{CDGT\ ,G<\tilde{G}_{cr}} = 0$. Hence the perturbations are zero, and the interface velocity for this stable solution is constant $\tilde{\mathbf{V}} = \tilde{\mathbf{V}}_0$.

We see that in the stable regime, $G < \tilde{G}_{cr}$ ($T > \tilde{T}_{cr}$), the resultant accelerated conservative dynamics with surface tension corresponds to the stable unperturbed flow fields $(\rho, \mathbf{V}, P_0, W_0)_{h(l)}$ and has constant interface velocity $\tilde{\mathbf{V}} = \tilde{\mathbf{V}}_0$, Figures 4-6, Tables 1,5. In this case the buoyancy (the gravity) is dominated the combined effects of the inertial stabilization mechanism and the surface tension (the reactive force and the tension force). The inertial stabilization mechanism is ‘masked’ for $T \neq 0$, since the integration constant for the solution is zero, and is exhibited only at $T \equiv 0$.

**<u>Unstable dynamics:</u>** The accelerated conservative dynamics with surface tension $\mathbf{r}_{CDGT}$ is unstable for $G > \tilde{G}_{cr}$ ($T < \tilde{T}_{cr}$), Figures 4,5,7, Tables 1,5. In this regime, the eigenvalue $\omega_{CDGT}$ is real and positive, $\mathrm{Re}[\omega_{CDGT}] > 0$ and $\mathrm{Im}[\omega_{CDGT}] = 0$. The dynamics $\mathbf{r}_{CDGT}$ couples the interface perturbation with the vortical and potential fields of the velocity $\nabla\varphi_h$, $\nabla\varphi_l$, $\nabla \times \boldsymbol{\psi}_l$. The potential and vortical components of the fluid velocities achieve their extreme values near the interface, and, while increasing in time, decay away from the interface.

The vortical field for the unstable solution $\mathbf{r}_{CDGT}$ with $G > \tilde{G}_{cr}$ ($T < \tilde{T}_{cr}$) in Eqs.(20) has the following properties. The wavevector of the vortical field is $\tilde{k} = (k/R)\,\omega_{CDGT}$, and the length-scale of the vortical field is large in a broad range of parameters, $\left(\tilde{k}/k\right) << 1$. When surface tension value decreases, $T \to 0$, the strength of the vortical field decreases, leading to the potential velocity fields are in the fluids’ bulks, Figures 4,5,7, Tables 1,5 [21,22].

For the unstable accelerated conservative dynamics with surface tension, the buoyancy (the gravity) dominates the combined effects of the inertial stabilization mechanism and the surface tension (the reactive force and the tension force), Figures 5,7, Tables 1,5. For $G > \tilde{G}_{cr}$ ($T < \tilde{T}_{cr}$), the interface velocity of the solution $\mathbf{r}_{CDGT}$ increases with time, $\tilde{\mathbf{V}} = \tilde{\mathbf{V}}_0 + \tilde{\mathbf{v}}$ with $\tilde{\mathbf{v}}\mathbf{n}_0 \sim e^{|\omega_{CDGT}|\,t}$. The resultant flow is the superposition of two motions – the motion of the interface as whole with the growing interface velocity and the growth of the interface perturbations. The dynamics is shear free at the interface. When compared to the accelerated conservative dynamics free from surface tension, the surface tension influences the acceleration values at which the instability occurs, and also leads to the appearance of the vortical field in the bulk of the light fluid, Figures 5,7, Tables 1,5, [21,22].

**Summary:** The accelerated conservative dynamics with surface tension can be stable of unstable depending on the values of the acceleration, the surface tension and the density ratio. In the stable regime, the resultant dynamics has the stable unperturbed flow fields $(\rho, \mathbf{V}, P_0, W_0)_{h(l)}$ and the constant interface velocity $\tilde{\mathbf{V}} = \tilde{\mathbf{V}}_0$. In the unstable regime, the interface perturbations grow and so is the interface velocity. The dynamics couples the interface perturbation with the potential velocity field in the heavy fluid bulk and the potential and vortical components of the velocity field in the light fluid bulk, and is shear-free at the interface. The strength of the vortical field in the light fluid bulk depends on the surface tension; for zero surface tension, the velocity fields are potential in both fluids, Figures 4,5,6,7, Tables 1,5.

**Sub-Section 3.5.2 – Classical Landau's dynamics**

**Fundamental solution**: For the classical Landau's dynamics in the presence of acceleration and surface tension the solution is $\mathbf{r}_{LDGT} = \mathbf{r}_{LDGT}\left(\omega_{LDGT}, \tilde{\mathbf{e}}_{LDGT}\right)$ Figures 4,5,8, Tables 2,5:

$$\omega_{LDGT} = \frac{-R + \sqrt{R^3 + R^2 - R + \overline{G}\left(R^2 - 1\right)}}{R+1}, \quad \mathbf{e}_{LDGT} = \mathbf{e} = \left(\varphi, \tilde{\varphi}, 1, \psi\right)^{\mathrm{T}}; \quad \overline{G} = G - \frac{T\,R}{R-1}; \quad (21)$$
$$\tilde{\mathbf{V}} = \tilde{\mathbf{V}}_0 + \tilde{\mathbf{v}}, \quad \tilde{\mathbf{v}}\mathbf{n}_0 = -\left(\mathbf{u}_h\mathbf{n}_0 + \dot{\theta}\right)\Big|_{\theta=0} \equiv 0, \quad \tilde{\mathbf{V}} \equiv \tilde{\mathbf{V}}_0$$

where the modified acceleration $\overline{G}$ accounts for the contribution of the surface tension $T$. The quantities $\{\varphi, \tilde{\varphi}, \psi\}$ depend on the values of the density ratio, the acceleration and the surface tension $R, G, T$. For $T \to 0$ this solution is consistent with the solution for the accelerated Landau's dynamics free from surface tension Eqs.(19.2), Table 2 [20-22].

**Stability and instability of the fundamental solution:** In the classical Landau's dynamics the inertial stabilization mechanism is absent, due to the postulate of the constancy of the interface velocity. The dynamics can be stabilized by the surface tension. The solution $\mathbf{r}_{LDGT}$ is stable for $T > \overline{T}_{cr}$ ( $G < \overline{G}_{cr}$) and is unstable for $T < \overline{T}_{cr}$ ($G > \overline{G}_{cr}$). The condition $\omega_{LDGT} = 0$ defines the critical values $\overline{T}_{cr} = (G + R)(R - 1)/R$ and $\overline{G}_{cr} = (TR - R(R - 1))/R$, Figures 4,5, Tables 2,5.

**Structure of flow fields:** In either stable or unstable regime, the dynamics $\mathbf{r}_{LDGT}$ couples the interface perturbation with the vortical and potential fields of the velocity $\nabla\varphi_h, \nabla\varphi_l, \nabla\times\boldsymbol{\psi}_l$. The presence of the vortical field in the classical Landau's dynamics is caused by the energy imbalance, which is due to the postulated constancy of the interface velocity $\tilde{\mathbf{V}} \equiv \tilde{\mathbf{V}}_0$ and the associated interfacial boundary condition for the perturbed velocity [21-23].

**Stable dynamics:** For $T > \overline{T}_{cr}$ ($G < \overline{G}_{cr}$), the solution is stable, with $\mathrm{Re}[\omega_{LDGT}] < 0$, and the length-scale of the vortical field $\tilde{k} = (k/R)\,\omega_{LDGT}$ has the negative real part $\mathrm{Re}\,[\tilde{k}\,] < 0$, Figures 4,5, Tables 2,5. The vortical component of the velocity of the light fluid $\nabla\times\boldsymbol{\psi}_l$ and its vorticity $\nabla\times\mathbf{u}_l = \left(0, \left(1 - (\tilde{k}/k)^2\right)\psi_l, 0\right)$ increase far from the interface, $z \to +\infty$. In order for the solution $\mathbf{r}_{LDGT}$ to obey the boundary condition $\mathbf{u}_l\big|_{z\to+\infty} = 0$ in Eqs.(8.2,12.1), we must set its integration constant equal zero $C_{LDGT} = 0$. Hence, in the stable regime, $T > \overline{T}_{cr}$ ($G < \overline{G}_{cr}$), the Landau's dynamics with the acceleration and the surface tension has unperturbed flow fields $(\rho, \mathbf{V}, P_0, W_0)_{h(l)}$ and constant interface velocity $\tilde{\mathbf{V}} = \tilde{\mathbf{V}}_0$.

**Unstable dynamics:** For $T < \overline{T}_{cr}$ ($G > \overline{G}_{cr}$) the accelerated Landau's dynamics with surface tension $\mathbf{r}_{LDGT}$ is unstable, Eqs.(21), Figures 4,5,8, Tables 2,5. In this regime, the eigenvalue $\omega_{LDGT}$ is real and positive, $\mathrm{Re}[\omega_{LDGT}] > 0$ and $\mathrm{Im}[\omega_{LDGT}] = 0$. The dynamics $\mathbf{r}_{LDGT}$ couples the interface perturbation with the vortical and potential fields of the velocity $\nabla\varphi_h, \nabla\varphi_l, \nabla\times\boldsymbol{\psi}_l$. The potential and vortical components of the fluid velocities achieve their maximum values near the interface, and, while increasing in time, decay away from the interface, Figure 8. For the unstable solution $\mathbf{r}_{LDGT}$ with $T < \overline{T}_{cr}$ in Eqs.(21), the vortical field has the wavevector $\tilde{k} = (k/R)\,\omega_{LDGT}$; the length-scale of the vortical field is large, $(\tilde{k}/k) << 1$, in a broad range of parameters. While the vortical field depends on the surface tension value, it is present for any values of the acceleration and the surface tension, and is associated with the energy imbalance for the Landau's dynamics [21-23]. Hence, in the unstable regime, $T < \overline{T}_{cr}$ ($G > \overline{G}_{cr}$) the accelerated Landau's dynamics with surface tension is the superposition of two

motions – the motion of the interface with the constant velocity $\tilde{\mathbf{V}} \equiv \tilde{\mathbf{V}}_0$ and the growth of the interface perturbations. It is shear free at the interface.

**Summary:** The accelerated Landau's dynamics with surface tension can be stable or unstable depending on the values of the acceleration, the surface tension and the density ratio. In the stable regime, the resultant dynamics corresponds to the stable unperturbed flow fields $(\rho, \mathbf{V}, P_0, W_0)_{h(l)}$ and has constant interface velocity $\tilde{\mathbf{V}} \equiv \tilde{\mathbf{V}}_0$. In the unstable regime, the interface perturbations grow, whereas the interface velocity remains constant. The unstable dynamics couples the interface perturbation with the potential and vortical components of the velocity fields in the fluids' bulk, and is shear-free at the interface. The presence of the vortical field in the light fluid bulk is due to the postulated constancy of the interface velocity, leading to energy imbalance for any value of the acceleration and the surface tension, Figures 4,5,8, Tables 2,5 [21-23].

**Sub-Section 3.5.3 – Rayleigh-Taylor dynamics**

**Fundamental solution**: For Rayleigh-Taylor dynamics in the presence of acceleration and surface tension the solution is $\mathbf{r}_{RDGT} = \mathbf{r}_{RDGT}(\omega_{RDGT}, \tilde{\mathbf{e}}_{RDGT})$, Figures 4,5,9, Tables 3,5

$$\overline{G} < 0, \quad \omega_{RTGT} = \pm i\sqrt{\overline{G}}\sqrt{\frac{R-1}{R+1}}, \quad \mathbf{e}_{RTGT} = \frac{\mathbf{e} + \mathbf{e}^*}{2}, \quad \mathbf{e} = (\varphi, \tilde{\varphi}, 1)^{\mathrm{T}}; \quad (22)$$
$$\overline{G} > 0, \quad \omega_{RTGT} = \sqrt{\overline{G}}\sqrt{\frac{R-1}{R+1}}, \quad \mathbf{e}_{RTGT} = \mathbf{e} = (\varphi, \tilde{\varphi}, 1)^{\mathrm{T}}; \quad \overline{G} = G - \frac{T\,R}{R-1}$$

where the components of eigenvectors $\{\varphi, \tilde{\varphi}\}$ for solutions are the functions on $R, G, T$, Table 3.

**Stability and instability of the fundamental solution:** Rayleigh-Taylor dynamics is stabilized by the surface tension. The solution $\mathbf{r}_{RTGT}$ is stable for $T > \hat{T}_{cr}$ and is unstable for $0 < T < \hat{T}_{cr}$. The critical surface tension value $\hat{T}_{cr} = G(R-1)/R$ is defined by the condition $\overline{G} = 0$, Tables 3,5; for $G \to 0$ the value approaches $\hat{T}_{cr} \to 0$, in agreement with Eqs.(19.3).

**Structure of flow fields** In the stable and the unstable regimes of Rayleigh-Taylor dynamics with surface tension, the velocity fields are potential in the bulks of the light and the heavy fluid, and have the interfacial shear at the interface, Figure 9.

**Summary**: For Rayleigh-Taylor dynamics with surface tension in the stable regime the dynamics describes the standing wave stably oscillating in time (which is the capillary wave for zero acceleration). In the unstable regime the dynamics describes the standing wave with the growing amplitude. The velocity fields are potential in the bulks of the light and the heavy fluid, and there is the shear at the interface, Figure 9, Table 3 [20-31]. The interface velocity is zero in the laboratory reference frame.

**Sub-Section 3.5.4 – Properties of the accelerated interfacial dynamics with surface tension**

Depending on the values of the acceleration, the surface tension and the density ratio, the dynamics $\{\mathbf{r}_{CDGT}, \mathbf{r}_{LDGT}, \mathbf{r}_{RTGT}\}$ can be stable or unstable, Figures 4-9, Tables 1-3,5,6. In either stable or unstable regime, these dynamics have distinct qualitative properties.

The accelerated conservative dynamics is stable (unstable) for $T > (<)\, \tilde{T}_{cr}$ and $G < (>)\, \tilde{G}_{cr}$. In the stable regime, the resultant motion corresponds to unperturbed flow fields with constant interface velocity and zero interfacial shear. In the unstable regime, the dynamics couples the interface perturbations with the potential and vortical components of the velocity fields in the fluids' bulks and is shear free at the interface; the interface perturbations grow with time and so is the interface velocity, Figures 4-7, Tables 1,5,6.

The accelerated Landau's dynamics is stable (unstable) for $T > (<)\, \overline{T}_{cr}$ and $G < (>)\, \overline{G}_{cr}$. In the stable regime, the resultant dynamics corresponds to unperturbed flow fields with constant interface velocity and with zero interfacial shear. In the unstable regime, the dynamics couples the interface perturbations with the potential and vortical components of the velocity fields in the fluids' bulks and is shear free at the interface; the interface perturbations grow with time; the interface velocity is constant, Figures 4,5,8, Tables 2,5,6.

Rayleigh-Taylor dynamics is stable (unstable) for $T > (<)\, \hat{T}_{cr}$ and $G < (>)\, \hat{G}_{cr}$. In either regime the dynamics has potential velocity fields in the fluid bulks and has the interfacial shear; the interface velocity is zero in the laboratory frame of reference. In the stable regime, the dynamics describes the stably oscillating standing wave. In the unstable regime, the amplitude of the standing wave grows with time, Figures 4,5,9, Tables 3,5,6.

**Sub-Section 3.6 – Mechanisms of stabilization and destabilization**

By comparing properties of fundamental solutions $\{\mathbf{r}_{CDGT}, \mathbf{r}_{LDGT}, \mathbf{r}_{RTGT}\}$ we further analyze the mechanisms of stabilization and destabilization of the interface dynamics influenced by the acceleration and surface tension, Figures 4-9, Tables 1-3,5,6.

**Sub-Section 3.6.1 – Acceleration**

Since the acceleration is directed from the heavy fluid to the light fluid, its qualitative role is to destabilize the interface dynamics. Quantitative effect of the acceleration is however distinct for the conservative, Landau's and Rayleigh-Taylor dynamics.

By comparing the growth-rates' values for the dynamics $\{\mathbf{r}_{CDGT}, \mathbf{r}_{LDGT}, \mathbf{r}_{RTGT}\}$, we find that at $T=0$ they are $\omega_{CDGT} = \omega_{LDGT} = \omega_{RTGT}$ at $G = G^* = (R^2-1)/4$. For strong accelerations and weak surface tension the dynamics $\{\mathbf{r}_{CDGT}, \mathbf{r}_{LDGT}, \mathbf{r}_{RTGT}\}$ are unstable, and the growth-rates behave as

$$G \to \infty, \quad T \to 0: \qquad \omega_{CDGT} > \omega_{LDGT} > \omega_{RTGT} \tag{23}$$

$$\omega_{CDGT} \to \sqrt{G}\sqrt{\frac{R+1}{R-1}} + \frac{1}{2\sqrt{G}}\left(-\frac{TR}{\sqrt{R^2-1}} - R\sqrt{\frac{R-1}{R+1}}\right);$$

$$\omega_{LDGT} \to \sqrt{G}\sqrt{\frac{1}{R+1}} + \left(-1 + \frac{1}{R+1}\right) + \frac{1}{2\sqrt{G}}\left(-\frac{TR}{\sqrt{R^2-1}} + \frac{R(R^2+R-1)}{(R+1)\sqrt{R^2-1}}\right);$$

$$\omega_{RTGT} \to \sqrt{G}\sqrt{\frac{R-1}{R+1}} + \frac{1}{2\sqrt{G}}\left(-\frac{TR}{\sqrt{R^2-1}}\right)$$

Hence, in the limit of strong accelerations and weak surface tension values, the new fluid instability of the conservative dynamics has the largest growth-rate when compared to the accelerated Landau's and Rayleigh-Taylor dynamics, Figures 5, Tables 1-3.

**Sub-Section 3.6.2 – Surface tension**

The surface tension qualitative role is to stabilize the interface dynamics. Quantitative effect of the surface tension is however distinct for the conservative, Landau's and Rayleigh-Taylor dynamics.

For given values of the acceleration and density ratio $G>0, R>1$, each of the dynamics $\{\mathbf{r}_{CDGT}, \mathbf{r}_{LDGT}, \mathbf{r}_{RTGT}\}$ can be stabilized by surface tension. The new fluid instability of the accelerated conservative dynamics is stabilized for $T > \tilde{T}_{cr}$. The Landau's dynamics is stabilized for $T > \overline{T}_{cr}$. Rayleigh-Taylor dynamics is stabilized for $T > \hat{T}_{cr}$. By comparing the critical surface tension values for given values of the acceleration and the density ratio, we find (Figures 4,5, Tables 1-3,5): For $G>0, R>1$ the values related as $\overline{T}_{cr} > \hat{T}_{cr}$, and the Landau's dynamics can be stabilized by larger surface tension when compared to Rayleigh-Taylor dynamics. For $G > R(R-1)/2$ the values relate as $\hat{T}_{cr} > \tilde{T}_{cr}$, and Rayleigh-Taylor dynamics can be stabilized by larger surface tension when compared to the new fluid instability of the conservative dynamics. For $G > R(R-1)$ the values relate as $\tilde{T}_{cr} > \hat{T}_{cr}$, and the Landau's dynamics can be stabilized by smaller surface tension when compared to the new fluid instability of the conservative dynamics, Figures 4,5, Table 5.

Hence, for weak accelerations, $G < R(R-1)/2$, the critical surface tension values relate as $\tilde{T}_{cr} < \hat{T}_{cr} < \overline{T}_{cr}$ and the new fluid instability of the accelerated dynamics is stabilized by the smallest surface tension value when compare to Rayleigh-Taylor and Landau's dynamics. For intermediate accelerations, $R(R-1)/2 < G < R(R-1)$ the values relate as $\hat{T}_{cr} < \tilde{T}_{cr} < \overline{T}_{cr}$. For strong accelerations, $G > R(R-1)$, the values relate as $\hat{T}_{cr} < \overline{T}_{cr} < \tilde{T}_{cr}$ and the new fluid instability of the conservative dynamics requires the largest surface tension value for the stabilization, when compared to Rayleigh-Taylor and Landau's dynamics, Figures 4,5, Table 5.

**Sub-Section 3.6.3 – Inertial stabilization mechanism**

The foregoing results have clear physics interpretation: The conservative dynamics has inertial stabilization mechanism. Hence, for weak accelerations, the presence of this mechanism leads to smaller values of surface tension required for the interface stabilization, when compared to Landau's and Rayleigh-Taylor dynamics. For strong accelerations, the conservative dynamics has the largest growth-rate leading to the largest surface tension value required for the interface stabilization, when compared to Landau's and Rayleigh-Taylor dynamics.

The inertial stabilization mechanism is the new mechanism recently discovered for the interface dynamics with interfacial mass flux [21,22]. This mechanism is the essential property of the dynamics at macroscopic continuous scales. It is associated with the conservation of momentum and energy in the fluid system [21,22]. It is exhibited in the non-constancy of the interface velocity for the unsteady non-planar interface. This mechanism is absent in the classical Landau's dynamics due to the postulated constancy of the interface velocity. It is also absent in Rayleigh-Taylor dynamics, in which the interface velocity is zero in the laboratory frame of reference.

In the stable regime, the inertial stabilization mechanism is revealed in slight oscillations of the interface velocity at zero surface tension. In the unstable regime, its presence is exhibited in the non-constancy of the interface velocity.

**Sub-Section 3.7 – Characteristic length scales**

The values of gravity $g$, the velocity $V_h$, the surface tension $\sigma$ and the fluid densities $\rho_{h(l)}$ define the characteristic length-scales and time-scales of the dynamics of ideal incompressible fluids. These include the critical value of the wavevector $k_{cr}$ at which the interface is stabilized, and the maximum value of the wavevector $k_{max}$ at which the maximum value is achieved of the growth-rate of the interface perturbations, and the associated length-scales (wavelengths) $\lambda_{cr(max)} = 2\pi/k_{cr(max)}$ and time-scales $\tau_{cr(max)} = \left(k_{cr(max)}V_h\right)^{-1}$. For given values of $V_h, g, \sigma, \rho_h, \rho_l$, we present in the dimensional form each of the dynamics $\{\mathbf{r}_{CDGT}, \mathbf{r}_{LDGT}, \mathbf{r}_{RTGT}\}$ and find the critical and the maximum wavevector values from the conditions (Tables 7,8):

$$\Omega\big|_{k=k_{cr}} = 0; \qquad \left.\frac{\partial\Omega}{\partial k}\right|_{k=k_{max}} = 0, \quad \left.\frac{\partial^2\Omega}{\partial k^2}\right|_{k=k_{max}} < 0; \qquad \Omega = \Omega_{CDGT(LDGT)(RTGT)} \qquad (24)$$

**<u>Accelerated conservative dynamics</u>**: For the fundamental solution $\mathbf{r}_{CDGT}$ with $\Omega = \Omega_{CDGT}$ in Eqs.(24), the critical and the maximum wavevector values are (Tables 7,8):

$$\Omega_{CDGT}\big|_{k=\tilde{k}_{kr}} = 0; \quad \left.\frac{\partial\Omega_{CDGT}}{\partial k}\right|_{k=\tilde{k}_{max}} = 0, \quad \left.\frac{\partial^2\Omega_{CDGT}}{\partial k^2}\right|_{k=\tilde{k}_{max}} < 0 \quad (25.1)$$

$$\tilde{k}_{cr} = \frac{1}{2\sigma}\left[-V_h^2\left(\frac{\rho_h}{\rho_l}\right)(\rho_h-\rho_l) + \sqrt{4\sigma g(\rho_h+\rho_l) + \left(V_h^2\left(\frac{\rho_h}{\rho_l}\right)(\rho_h-\rho_l)\right)^2}\right]$$

$$\tilde{k}_{max} = \frac{1}{6\sigma}\left[-2V_h^2\left(\frac{\rho_h}{\rho_l}\right)(\rho_h-\rho_l) + \sqrt{12\sigma g(\rho_h+\rho_l) + \left(2V_h^2\left(\frac{\rho_h}{\rho_l}\right)(\rho_h-\rho_l)\right)^2}\right]$$

For $\left|\sigma g(\rho_h+\rho_l)/\left(V_h^2(\rho_h/\rho_l)(\rho_h-\rho_l)\right)^2\right| \to 0$ the values $\tilde{k}_{cr(max)}/\left(g/V_h^2\right) \sim O(1)$ remain finite, whereas for $\left|\sigma g(\rho_h+\rho_l)/\left(V_h^2(\rho_h/\rho_l)(\rho_h-\rho_l)\right)^2\right| \to \infty$ the values approach $\tilde{k}_{cr(max)}/\left(g/V_h^2\right) \to 0$.

The ratio $\left(\tilde{k}_{cr}/\tilde{k}_{max}\right)$ is the function on the parameters $V_h, g, \sigma, \rho_h, \rho_l$, Tables 7,8. For vanishing surface tension, the critical and maximum wave-vector values and their ratio are:

$$\left|\frac{\sigma}{V_h^2(\rho_h-\rho_l)}\frac{g}{V_h^2}\left(\frac{\rho_h+\rho_l}{\rho_h-\rho_l}\right)\left(\frac{\rho_h}{\rho_l}\right)^2\right| \to 0; \quad (25.2)$$

$$\tilde{k}_{cr} \to \frac{g}{V_h^2}\frac{\rho_l}{\rho_h}\left(\frac{\rho_h+\rho_l}{\rho_h-\rho_l}\right)\cdots; \quad \tilde{k}_{max} \to \frac{1}{2}\frac{g}{V_h^2}\frac{\rho_l}{\rho_h}\left(\frac{\rho_h+\rho_l}{\rho_h-\rho_l}\right); \quad \frac{\tilde{k}_{cr}}{\tilde{k}_{max}} \to 2$$

For very large surface tension, the critical and maximum wave-vector values and their ratio are:

$$\left|\frac{\sigma}{V_h^2(\rho_h-\rho_l)}\frac{g}{V_h^2}\left(\frac{\rho_h+\rho_l}{\rho_h-\rho_l}\right)\left(\frac{\rho_h}{\rho_l}\right)^2\right|\to\infty \quad (25.3)$$

$$\tilde{k}_{cr}\to\sqrt{\frac{g(\rho_h+\rho_l)}{\sigma}};\quad \tilde{k}_{max}\to\frac{1}{\sqrt{3}}\sqrt{\frac{g(\rho_h+\rho_l)}{\sigma}};\quad \frac{\tilde{k}_{cr}}{\tilde{k}_{max}}\to\sqrt{3}$$

**<u>Landau's dynamics</u>**: For the fundamental solution $\mathbf{r}_{LDGT}$ with $\Omega=\Omega_{LDGT}$ Eqs.(24), the critical and maximum wavevector values are (Tables 7,8):

$$\Omega_{LDGT}\big|_{k=\bar{k}_{kr}}=0;\quad \frac{\partial\Omega_{LDGT}}{\partial k}\bigg|_{k=\bar{k}_{max}}=0,\quad \frac{\partial^2\Omega_{LDGT}}{\partial k^2}\bigg|_{k=\bar{k}_{max}}<0 \quad (26.1)$$

$$\bar{k}_{cr}=\frac{1}{2\sigma}\left[-V_h^2\left(\frac{\rho_h}{\rho_l}\right)(\rho_h-\rho_l)+\sqrt{4\sigma g(\rho_h-\rho_l)+\left(V_h^2\left(\frac{\rho_h}{\rho_l}\right)(\rho_h-\rho_l)\right)^2}\right]$$

$$\bar{k}_{max}=\bar{k}_{max}(V_h,g,\sigma,\rho_h,\rho_l)$$

For vanishing surface tension values $\left|\sigma g/\left(V_h^4(\rho_h/\rho_l)^2(\rho_h-\rho_l)\right)\right|\to 0$ the critical and maximum wave-vectors values approach $\bar{k}_{cr(max)}/\left(g/V_h^2\right)\to\infty$, whereas for $\left|\sigma g/\left(V_h^4(\rho_h/\rho_l)^2(\rho_h-\rho_l)\right)\right|\to\infty$, the values are $\bar{k}_{cr(max)}/\left(g/V_h^2\right)\to 0$.

The ratio $\left(\bar{k}_{cr}/\bar{k}_{max}\right)$ is a cumbersome function on the parameters $V_h,g,\sigma,\rho_h,\rho_l$, Tables 7,8. For vanishing surface tension, the critical and maximum wave-vector values and their ratio are:

$$\left|\frac{\sigma}{V_h^2(\rho_h-\rho_l)}\frac{g}{V_h^2}\left(\frac{\rho_h}{\rho_l}\right)^2\right|\to 0;\qquad \bar{k}_{cr}\to\frac{\rho_h V_h^2}{\sigma}\left(\frac{\rho_h}{\rho_l}\right)\left(1-\frac{\rho_l}{\rho_h}\right);\quad (26.2)$$

$$\bar{k}_{max}\to\frac{\rho_h V_h^2}{\sigma}s\left(\frac{\rho_h}{\rho_l}\right),\quad s\big|_{\frac{\rho_h}{\rho_l}\to 1^+}\to\frac{1}{2}\left(1-\left(\frac{\rho_l}{\rho_h}\right)\right),\quad s\big|_{\frac{\rho_h}{\rho_l}\to\infty}\to\frac{2}{3}\left(\frac{\rho_h}{\rho_l}\right);$$

$$\frac{\bar{k}_{cr}}{\bar{k}_{max}}\bigg|_{\frac{\rho_h}{\rho_l}\to 1^+}\to 2,\qquad \frac{\bar{k}_{cr}}{\bar{k}_{max}}\bigg|_{\frac{\rho_h}{\rho_l}\to\infty}\to\frac{3}{2}$$

For very large surface tension, the critical and maximum wave-vector values and their ratio are:

$$\left|\frac{\sigma}{V_h^2(\rho_h-\rho_l)}\frac{g}{V_h^2}\left(\frac{\rho_h}{\rho_l}\right)^2\right|\to\infty;\qquad \bar{k}_{cr}\to\sqrt{\frac{g(\rho_h-\rho_l)}{\sigma}};\quad (26.3)$$

$$\bar{k}_{max}\to\sqrt{\frac{g(\rho_h-\rho_l)}{\sigma}}p\left(\frac{\rho_h}{\rho_l}\right),\quad p\big|_{\frac{\rho_h}{\rho_l}\to 1^+}\to\frac{1}{\sqrt{3}}\sqrt{\frac{\rho_h}{\rho_l}-1},\quad p\big|_{\frac{\rho_h}{\rho_l}\to\infty}\to\frac{1}{\sqrt{3}}$$

$$\frac{\bar{k}_{cr}}{\bar{k}_{max}}\bigg|_{\frac{\rho_h}{\rho_l}\to 1^+}\to\sqrt{3},\qquad \frac{\bar{k}_{cr}}{\bar{k}_{max}}\bigg|_{\frac{\rho_h}{\rho_l}\to\infty}\to\sqrt{3}$$

**Rayleigh-Taylor dynamics**: For the fundamental solution $\mathbf{r}_{RTGT}$ with $\Omega = \Omega_{RTGT}$ in Eqs.(24), the critical and the maximum wavevector values are (Tables 7,8):

$$\Omega_{RTGT}\big|_{k=\hat{k}_{cr}} = 0; \quad \left.\frac{\partial \Omega_{RTGT}}{\partial k}\right|_{k=\hat{k}_{max}} = 0, \quad \left.\frac{\partial^2 \Omega_{RTGT}}{\partial k^2}\right|_{k=\hat{k}_{max}} < 0 \quad (27)$$

$$\hat{k}_{cr} = \sqrt{\frac{g}{\sigma}(\rho_h - \rho_l)}; \quad \hat{k}_{max} = \frac{1}{\sqrt{3}}\sqrt{\frac{g}{\sigma}(\rho_h - \rho_l)}; \quad \frac{\hat{k}_{cr}}{\hat{k}_{max}} = \sqrt{3}$$

For vanishing surface tension values, $\left|\sigma/\left(g(\rho_h - \rho_l)\right)\right| \to 0$, the critical and maximum wavevector values approach $\hat{k}_{cr(max)}/\left(g/V_h^2\right) \to \infty$, whereas for very large surface tension values $\left|\sigma/\left(g(\rho_h - \rho_l)\right)\right| \to \infty$ the critical and maximum wavevector values approach $\hat{k}_{cr(max)}/\left(g/V_h^2\right) \to 0$ for given finite values of $g, V_h$, where $V_h$ is understood as some velocity scale. The ratio of the critical and maximum wavevector values is $\left(\hat{k}_{max}/\hat{k}_{cr}\right) = 1/\sqrt{3}$ for any $\left|\sigma/\left(g(\rho_h - \rho_l)\right)\right| > 0$, Tables 7,8.

**Comparative analysis:** The conservative dynamics of the fluid interface is stabilized by the inertial mechanism and by the surface tension, and is destabilized by the acceleration. The presence of the inertial stabilization mechanism is revealed in the finite values of the critical and the maximum wavevector values $\tilde{k}_{cr(max)}/\left(g/V_h^2\right) \sim O(1)$ in the limit of vanishing surface tension. For very large surface tension values the critical and the maximum wavevector values approach zero $\tilde{k}_{cr(max)}/\left(g/V_h^2\right) \to 0$. The ratio $\left(\tilde{k}_{cr}/\tilde{k}_{max}\right)$ is the function on the parameters $V_h, g, \sigma, \rho_h, \rho_l$, and it varies from $2$ to $\sqrt{3}$ with the increase of the surface tension parameter $\left|\sigma g(\rho_h + \rho_l)/\left(V_h^2(\rho_h/\rho_l)(\rho_h - \rho_l)\right)^2\right|$ from zero to infinity, Tables 7,8.

In the Landau's and Rayleigh-Taylor dynamics the properties of the characteristic scales are distinct when compared to those in the conservative dynamics, Tables 7,8. The Landau's dynamics is stabilized by surface tension, and it is unstable even for zero acceleration. For given values of $g, V_h$, in the limit of vanishing surface tension values the critical and maximum wavevector values approach $\bar{k}_{cr(max)}/\left(g/V_h^2\right) \to \infty$. For very large surface tension values, the critical and maximum wavevector values approach $\bar{k}_{cr(max)}/\left(g/V_h^2\right) \to 0$. The ratio $\left(\bar{k}_{cr}/\bar{k}_{max}\right)$ depends on the density ratio $(\rho_h/\rho_l)$ and the surface tension parameter $\left|\sigma g/\left(V_h^4(\rho_h/\rho_l)^2(\rho_h - \rho_l)\right)\right|$. With the increase of this parameter the ratio $\left(\bar{k}_{cr}/\bar{k}_{max}\right)$ varies from $2$ for $(\rho_h/\rho_l) \sim 1$ and $3/2$ for $(\rho_h/\rho_l) >> 1$ to $\sqrt{3}$ for any density ratio $(\rho_h/\rho_l)$, Tables 7,8.

Rayleigh-Taylor dynamics is stabilized by surface tension and is destabilized by the acceleration. For vanishing surface tension values, the critical and maximum wavevector values approach $\hat{k}_{cr(max)} / \left( g / V_h^2 \right) \to \infty$, whereas for very large surface tension values the critical and maximum wavevector values approach $\hat{k}_{cr(max)} / \left( g / V_h^2 \right) \to 0$, where $V_h$ is some velocity scale. The ratio of the critical and maximum wavevector values is $\left( \hat{k}_{cr} / \hat{k}_{max} \right) = \sqrt{3}$ for any value of the surface tension parameter $\left| \sigma / \left( g \left( \rho_h - \rho_l \right) \right) \right| > 0$, Tables 7,8.

We can conclude that the boundary conditions at the interface strongly influence the characteristic wave-vectors, length-scales and time-scales of the interfacial dynamics.

**Sub-Section 3.8 – Outcome for experiments and simulations**

Our analysis identifies the mechanisms of stabilization and destabilization of the interface dynamics with the interfacial mass flux and finds that the properties of the inertial and accelerated conservative dynamics with surface tension differ qualitatively and quantitatively from those of classical Landau's dynamics and Rayleigh-Taylor dynamics [21]. This opens new opportunities for experiments and simulations, and enables a better understanding and, ultimately, control of a broad range of processes in nature and technology to which unstable interfaces and interfacial mixing are relevant [1-45]. In this section we outline the outcomes of our analysis for experiments and simulations.

**<u>Outcomes for experiments and simulations</u>**: In order to compare with existing experiments and simulations, we note that our results for the Landau's dynamics and Rayleigh-Taylor dynamics agree with the results of theoretical, experimental and numerical studies [24-35]. Furthermore, our results for the conservative dynamics clearly indicate that the interface can be stable even for ideal incompressible fluids with vanishing surface tension, when the acceleration value is smaller than a threshold, similarly to ablative Rayleigh-Taylor instabilities in plasmas [36-40].

Our theory elaborates extensive benchmark for future experiments and simulations. According to our results, for given values of the fluid densities $\rho_{h(l)}$ and the velocity $V_h$, in the regime of strong accelerations $g$, the new fluid instability of the conservative dynamics has the largest stabilizing surface tension $\sigma$ and the largest growth-rate $\Omega$, when compared to the cases of the accelerated Landau's and Rayleigh-Taylor dynamics, Eqs.(11,12,14), Figures 5,7, Tables 1,5,6. The new fluid instability is the fastest in the extreme regimes of strong accelerations and weak surface tension, occurring, for instance in high energy density plasmas [5-9]. Hence, for given values of the parameters $V_h, \rho_{h(l)}, \sigma$, one can observe the new fluid instability by increasing the acceleration values. One can further observe that for the unstable accelerated conservative dynamics with surface tension, the growth of the interface perturbations is augmented with the growth of the interface velocity. The former is present and the latter is absent in the unstable regimes of the accelerated Landau's and Rayleigh-Taylor dynamics with surface tension Figure 7, Table 6 [24-35]. By accurately diagnosing the interface dynamics, including the growth of the interface perturbations and the interface velocity, one can confidently identify the new fluid instability in experiments with strong accelerations [5-9].

In some experiments the parameters of the dynamics $V_h, g, \sigma, \rho_h, \rho_l$ may be a challenge to vary systematically [5-9,39,40]. Our analysis proposes how to address the challenge. Particularly, for given values of the parameters $V_h, \rho_{h(l)}, \sigma, g$, by varying the wavelength of the initial perturbation $\lambda$, one may observe the interface stabilization at the wavevector $k = k_{cr}$ and the fastest growth-rate of the unstable interface at the wavevector $k = k_{max}$ and the associated length-scales $\lambda_{cr(max)} = 2\pi / k_{cr(max)}$ and time-scales $\tau_{cr(max)} = \left(k_{cr(max)} V_h\right)^{-1}$. One can further identify the type of the fluid instability, and differentiate between the new fluid instability of the conservative dynamics and the instabilities of the Landau's and Rayleigh-Taylor dynamics by comparing the critical and maximum scales $\lambda_{cr(max)}, \tau_{cr(max)}$ with the theoretical results for given values of the parameters, $V_h, \rho_{h(l)}, \sigma, g$, Eqs.(25-27), Tables 7,8. These results can be applied for design of experiment in high energy density plasmas [5-9,39,40].

Our results indicate a need in further advancements of numerical modeling of the interface dynamics [41-43]. Numerical modeling of unstable fluid interfaces is a challenge because the simulations are required to track the interface, to capture small scales dissipative processes, and to use the highly accurate numerical methods and massive computations [1]. Existing numerical approaches usually apply diffusive approximation for modeling interfaces with interfacial mass flux, and work well for flows with smoothly changing of flow fields [2]. New developments are in demand to accurately model the unstable interface with sharply changing flow fields, including the Lagrangian and Eulerian methods [1,2,41-43].

The existing experimental and numerical studies of the interface stability are focused on the measurements of the growth and growth-rate of the perturbation amplitude [2,5-7,39-43]. We derive the amplitude growth and the growth-rate, and we finds that the flow dynamics is highly sensitive to the interfacial boundary conditions, Figures 2-9, Tables 1-8. Our analysis directly links the macroscopic flow fields to the microscopic transport at the interface. It suggests that by measuring the flow fields at macroscopic scales in the bulk far from the interface, one can confidently capture the transport properties at microscopic scales at the interface, Figures 2-8. This information is especially important for systems where experimental data are a challenge to obtain, including fusion, supernovae and scramjets [2-15].

It is traditionally believed that the interface dynamics can be stabilized by factors depending on microscopic properties of matter (plasmas, fluids, materials), such as surface tension, diffusion, and dissipation, which, in turn, occur due to interactions of the constituting particles (atoms, molecules) [20,30-35]. Our analysis suggests that while these factors indeed play a stabilizing role, the conservative dynamics of the interface with interfacial mass flux can also be stabilized by the inertial mechanism, which is enabled by the macroscopic motion of the interface as whole Eqs.(11,20) [21,22]. This mechanism is absent in the classical Landau's dynamics, due to the postulated constancy of the interface velocity, and in Rayleigh-Taylor dynamics, where the interface velocity is zero in laboratory frame of reference Eqs.(12,14,21,22) [20-22]. In case of the accelerated dynamics with surface tension, the inertial mechanism is exhibited in the larger (smaller) values of surface tension required to stabilize the strongly (weakly) accelerated interface in the conservative dynamics when compared to the Landau's and Rayleigh-Taylor dynamics, Figures 3-9, Tables 1-5.

## Section 4 – Discussion and conclusion

We investigated the interfacial dynamics with interfacial mass flux in the presence of the acceleration and the surface tension Eqs.(1-27). We considered ideal and incompressible fluids with negligible stratification and densities variation for the two-dimensional spatially extended periodic flow with the acceleration directed from the heavy to the light fluid and with surface tension understood as the tension at the boundary between the flow phases. The general matrix method was advanced and applied to rigorously solve the linearized boundary value problem. The fundamental solutions were found for the dynamics conserving mass, momentum and energy, and were compared with those for the classical Landau's and Rayleigh-Taylor dynamics. The interplay of the acceleration, surface tension and inertial stabilization mechanism was scrupulously studied and its effect of the interface stability and on the properties of the new fluid instability of the conservative dynamics was identified. Extensive benchmarks were elaborated for future experiments and simulations and for better understanding of natural and technological processes, to which unstable interfaces are relevant, Eqs.(1-27), Figures 1-9, Tables 1-8.

We found that the dynamics conserving mass, momentum and energy can be stable or unstable depending on the acceleration and the surface tension. In the stable regime, the conservative dynamics corresponds to non-perturbed flow fields in the bulk, is shear-free at the interface and has the constant interface velocity. The instability can develop only in the presence of the acceleration and only when its magnitude exceeds a threshold, Eqs.(20), Figures 2-7, Tables 1,4,5,6. This threshold value reflects the contributions of the inertial stabilization mechanism and the surface tension and is finite for zero surface tension. In the unstable regime, the interface perturbations are coupled with the potential and vortical components of the velocity fields in the fluids' bulk; for zero surface tension, the velocity fields are potential. The dynamics is shear-free at the interface. It describes the standing wave with the growing amplitude, and has the growing interface velocity, Figure 6. Depending on the acceleration and the surface tension, the fluid instability of the conservative dynamics can grow faster or slower when compared to the accelerated Landau's and Rayleigh-Taylor dynamics; it has the largest growth-rate and the largest stabilizing surface tension value in the extreme regime of strong accelerations, Eqs.(20-22), Figures 4-9, Tables 2-6. We also found the critical and maximum values of the wavevector of the initial perturbation at which the conservative instability is stabilized and at which it has the largest growth-rate, Tables 7,8. These unique quantitative and qualitative properties of the instability of the conservative dynamics clearly distinct it from other fluid instabilities, and call for further investigations, Eqs.(1-27), Figures 1-9, Tables 1-8.

Our results agree qualitatively with available observations and indicate a strong need in further experimental and numerical studies of the interface dynamics, and in the development of new methods of numerical modeling and experimental diagnostics. Existing experimental and numerical studies of the interface dynamics are focused on diagnostics of the growth of the amplitude of the initial perturbation [1,2,39-43]. Our analysis provides the amplitude growth-rate in a broad range of parameters, determines the regions of the experimental parameter of the stable and unstable dynamics, identifies the structure of the flow field and links them to the boundary conditions at the interface Eqs.(1-27), Figures 1-9, Tables 1-8. Particularly, according to our results, by measuring at macroscopic scales the flow fields in the bulk, one can capture the transport properties at microscopic scales at the interface, Figures 6-9, Table 6.

Our results can be further connected to realistic environments in plasmas, fluids and material, in which the dynamics is usually accompanied by dissipation, diffusion, compressibility, radiation transport, stratification, and non-local forces [2-15]. Our general theoretical approach can be extended to systematically incorporate these effects, to analyze the interplay of the interface stability with the structure of flow fields, and to elaborate a unified theory framework for studies of interfacial dynamics in a broad range of processes including the ablative Rayleigh-Taylor instabilities in fusion plasmas, the dynamics of reactive and super-critical fluids, and the D'yakov-Kontorovich instability of the shock waves [1-45]. We address these studies to the future.

**Section 5 - Acknowledgements**

The authors thank for support the University of Western Australia, AUS (project grant 10101047); and the National Science Foundation, USA (award 1404449).

**Section 6 – Data availability**

The methods, the results and the data presented in this work are freely available to the readers in the paper and on the request from the authors.

**Section 7 –Author's contributions**

The authors contributed to the work as follows: SIA designed research; DVI, SIA performed research; DVI, SIA analyzed data; DVI, SIA discussed results; DVI, SIA wrote the paper.

## Section 8 –References

## Section 9 –Tables

Table 1: Fundamental solution for the conservative dynamics with the acceleration and the surface tension

| $\mathbf{r} = \mathbf{r}_{CDGT}$ , $\omega = \omega_{CDGT}$ , $\mathbf{e} = \mathbf{e}_{CDGT} = \left(\varphi, \tilde{\varphi}, \bar{z}, \psi\right)^{\mathrm{T}}_{CDGT}$ | |
|---|---|
| $\omega$ | $\frac{\sqrt{G+R+GR-R^2-RT}}{\sqrt{-1+R}}$ |
| $\varphi$ | $-\frac{1}{R+G(1+R)-R(R^2+T)}\big(\sqrt{-1+R}(T\sqrt{R+G(1+R)-R(R+T)} - G\big(\sqrt{-1+R}+\sqrt{R+G(1+R)-R(R+T)}\big) - R(\sqrt{-1+R}+\sqrt{R+G(1+R)-R(R+T)}) + R^2(\sqrt{-1+R}+\sqrt{R+G(1+R)-R(R+T)}))\big)$ |
| $\tilde{\varphi}$ | $\frac{1}{-G(1+R)+R(-1+R^2+T)}\big(G(-1+R)R - G\sqrt{-1+R}\sqrt{R+G(1+R)-R(R+T)} - (-1+R)R((-1+R)R+T-\sqrt{-1+R}\sqrt{R+G(1+R)-R(R+T)})\big)$ |
| $\bar{z}$ | $1$ |
| $\psi$ | $-\frac{i(-1+R)RT}{R+G(1+R)-R(R^2+T)}$ |

Table 2: Fundamental solution for the Landau's dynamics with the acceleration and the surface tension

| $\mathbf{r} = \mathbf{r}_{LDGT}$ , $\omega = \omega_{LDGT}$ , $\mathbf{e} = \mathbf{e}_{LDGT} = \left(\varphi, \tilde{\varphi}, \bar{z}, \psi\right)^{\mathrm{T}}_{LDGT}$ | |
|---|---|
| $\omega$ | $\frac{-R+\sqrt{-G-R+R^2+GR^2+R^3-RT-R^2T}}{1+R}$ |
| $\varphi$ | $-\frac{R-\sqrt{G(-1+R^2)+R(-1+R+R^2-(1+R)T)}}{1+R}$ |
| $\tilde{\varphi}$ | $\frac{1}{(1+R)(R(2+R)-\sqrt{G(-1+R^2)+R(-1+R+R^2-(1+R)T)})}\big(G-GR^2+R(T+\sqrt{G(-1+R^2)+R(-1+R+R^2-(1+R)T)})+R(-2+R+R^2+T-\sqrt{G(-1+R^2)+R(-1+R+R^2-(1+R)T)})\big)$ |
| $\bar{z}$ | $1$ |
| $\psi$ | $\frac{iR(-1+R(2+R)-2\sqrt{G(-1+R^2)+R(-1+R+R^2-(1+R)T)})}{R(2+R)-\sqrt{G(-1+R^2)+R(-1+R+R^2-(1+R)T)}}$ |

Table 3: Fundamental solution for Rayleigh-Taylor with the acceleration and the surface tension

| $\mathbf{r} = \mathbf{r}_{RTGT}$ , $\omega = \omega_{RTGT}$ , $\mathbf{e} = \mathbf{e}_{RTGT} = (\varphi, \tilde{\varphi}, \bar{z})^{\mathrm{T}}_{RTGT}$ | |
|---|---|
| $\omega$ | $\frac{\sqrt{-1+R}\sqrt{-G+GR-RT}}{\sqrt{-1+R^2}}$ |
| $\varphi$ | $\frac{\sqrt{-1+R}\sqrt{G(-1+R)-RT}}{\sqrt{-1+R^2}}$ |
| $\tilde{\varphi}$ | $-\frac{\sqrt{-1+R}\sqrt{G(-1+R)-RT}}{\sqrt{-1+R^2}}$ |
| $\bar{z}$ | 1 |

Table 4: Regions of stability and instability for the inertial acceleration-free dynamics with the surface tension for the conservative, Landau's and Rayleigh-Taylor dynamics

| Dynamics | Stability region | Instability region | Critical value |
|---|---|---|---|
| $\mathbf{r}_{CDGT}\big\|_{G=0}$ | $T > \tilde{T}_{cr}\big\|_{G=0}$ | $N/A$ | $\tilde{T}_{cr}\big\|_{G=0} = 0$ |
| $\mathbf{r}_{LDGT}\big\|_{G=0}$ | $T > \overline{T}_{cr}\big\|_{G=0}$ | $T < \overline{T}_{cr}\big\|_{G=0}$ | $\overline{T}_{cr}\big\|_{G=0} = (R-1)$ |
| $\mathbf{r}_{RTGT}\big\|_{G=0}$ | $T > \hat{T}_{cr}\big\|_{G=0}$ | $N/A$ | $\hat{T}_{cr}\big\|_{G=0} = 0$ |

Table 5: Regions of stability and instability and critical parameters for the accelerated dynamics with the surface tension for the conservative, Landau's and Rayleigh-Taylor dynamics

| Dynamics | Stability region | Instability region | Critical values |
|---|---|---|---|
| $\mathbf{r}_{CDGT}$ | $T > \tilde{T}_{cr}$ | $T < \tilde{T}_{cr}$ | $\tilde{T}_{cr} = \frac{G(R+1)-R(R-1)}{R}$ |
| | $G < \tilde{G}_{cr}$ | $G > \tilde{G}_{cr}$ | $\tilde{G}_{cr} = R\frac{R-1}{R+1} + T\frac{R}{R+1}$ |
| $\mathbf{r}_{LDGT}$ | $T > \overline{T}_{cr}$ | $T < \overline{T}_{cr}$ | $\overline{T}_{cr} = \frac{(G+R)(R-1)}{R}$ |
| | $G < \overline{G}_{cr}$ | $G > \overline{G}_{cr}$ | $\overline{G}_{cr} = T\frac{R}{R-1} - R$ |
| $\mathbf{r}_{RTGT}$ | $T > \hat{T}_{cr}$ | $T < \hat{T}_{cr}$ | $\hat{T}_{cr} = G\frac{(R-1)}{R}$ |
| | $G < \hat{G}_{cr}$ | $G > \hat{G}_{cr}$ | $\hat{G}_{cr} = T\frac{R}{R-1}$ |

Table 6: Qualitative properties of the conservative, Landau's and Rayleigh-Taylor dynamics with the acceleration and surface tension in their corresponding stable and unstable regimes

| Dynamics | Regime | Interface velocity | Velocity fields | Interfacial shear |
|---|---|---|---|---|
| $\mathbf{r}_{CDGT}$ | Stable | Constant | Unperturbed fields | Shear-free |
| | Unstable | Time-dependent | Potential and vortical components | Shear-free |
| $\mathbf{r}_{LDGT}$ | Stable | Constant | Unperturbed fields | Shear-free |
| | Unstable | Constant | Potential and vortical components | Shear-free |
| $\mathbf{r}_{RTGT}$ | Stable | Zero | Potential fields | Interfacial shear |
| | Unstable | Zero | Potential fields | Interfacial shear |

Table 7: Value of critical wavevector for the conservative, Landau's and Rayleigh-Taylor dynamics

| | | |
|---|---|---|
| $\mathbf{r}_{CDGT}$ | $\tilde{k}_{cr}$ | $\frac{1}{2\sigma}\left[-V_h^2\left(\frac{\rho_h}{\rho_l}\right)(\rho_h-\rho_l)+\sqrt{4\sigma g(\rho_h+\rho_l)+\left(V_h^2\left(\frac{\rho_h}{\rho_l}\right)(\rho_h-\rho_l)\right)^2}\right]$ |
| $\mathbf{r}_{LDGT}$ | $\bar{k}_{cr}$ | $\frac{1}{2\sigma}\left[-V_h^2\left(\frac{\rho_h}{\rho_l}\right)(\rho_h-\rho_l)+\sqrt{4\sigma g(\rho_h-\rho_l)+\left(V_h^2\left(\frac{\rho_h}{\rho_l}\right)(\rho_h-\rho_l)\right)^2}\right]$ |
| $\mathbf{r}_{RTGT}$ | $\hat{k}_{cr}$ | $\sqrt{\frac{g}{\sigma}(\rho_h-\rho_l)}$ |

Table 8: Value of maximum wavevector for the conservative, Landau's and Rayleigh-Taylor dynamics

| | | |
|---|---|---|
| $\mathbf{r}_{CDGT}$ | $\tilde{k}_{max}$ | $\frac{1}{6\sigma}\left[-2V_h^2\left(\frac{\rho_h}{\rho_l}\right)(\rho_h-\rho_l)+\sqrt{12\sigma g(\rho_h+\rho_l)+\left(2V_h^2\left(\frac{\rho_h}{\rho_l}\right)(\rho_h-\rho_l)\right)^2}\right]$ |
| $\mathbf{r}_{LDGT}$ | $\bar{k}_{max}$ | $\bar{k}_{max}=\bar{k}_{max}(V_h, g, \sigma, \rho_h, \rho_l)$ |
| $\mathbf{r}_{RTGT}$ | $\hat{k}_{max}$ | $\frac{1}{\sqrt{3}}\sqrt{\frac{g}{\sigma}(\rho_h-\rho_l)}$ |

## Section 10 – Figure captions and Figures

Figure 1: Schematics of the dynamics in a far field (not to scale). Blue color marks the planar (dashed line) interface and the perturbed (solid line) interface.

Figure 2: Growth-rates / frequencies for the inertial conservative dynamics (purple), Landau's dynamics (blue) and Rayleigh-Taylor dynamics (light blue) with the surface tension and free from the acceleration. Dependence of the growth-rates: (top) on the surface tension at some value of the density ratio; (bottom) on the density ratio at some value of the surface tension. Solid (dashed) line marks real (imagine) part.

Figure 3: Flow fields for the inertial conservative dynamics with surface tension at some instance of time and at some values of the density ratio and the surface tension: (a) plots of the perturbed velocity vector fields, the perturbed velocity streamlines, and the interface perturbation; (b) plots of the vortical component of the perturbed velocity and the perturbed vorticity. Real parts of fields and functions are shown. Each plot has its own range of values to better present the plot's features.

Figure 4: Growth-rates / frequencies for the accelerated conservative dynamics (purple), Landau's dynamics (blue) and Rayleigh-Taylor dynamics (light blue) with the surface tension at some values of the density ratio and the acceleration: Dependence of the growth-rates on the surface tension for (top) weak and (bottom) intermediate accelerations. Solid (dashed) line marks real (imagine) part.

Figure 5: Growth-rates / frequencies for the accelerated conservative dynamics (purple), Landau's dynamics (blue) and Rayleigh-Taylor dynamics (light blue) with the surface tension at some value of the density ratio. Dependence of the growth-rates: (top) on the surface tension at some acceleration value for strong accelerations; (bottom) on the acceleration at some values of the surface tension. Solid (dashed) line marks real (imagine) part.

Figure 6: Flow fields for the fundamental solution for the accelerated conservative dynamics with surface tension in the stable regime at some instance of time and at some values of the density ratio, the acceleration, and the surface tension: (a) plots of the perturbed velocity vector fields, the perturbed velocity streamlines, and the interface perturbation; (b) plots of the vortical component of the perturbed velocity and the perturbed vorticity. Real parts of fields and functions are shown. Each plot has its own range of values to better present the plot's features.

Figure 7: Flow fields for the fundamental solution for the accelerated Landau's dynamics with surface tension in the unstable regime at some instance of time and at some values of the density ratio, the acceleration, and the surface tension: (a) plots of the perturbed velocity vector fields, the perturbed velocity streamlines, and the interface perturbation; (b) plots of the vortical component of the perturbed velocity and the perturbed vorticity. Real parts of fields and functions are shown. Each plot has its own range of values to better present the plot's features.

Figure 8: Flow fields for the fundamental solution for Rayleigh-Taylor dynamics with surface tension at some instance of time and at some values of the density ratio, the acceleration, and the surface tension: plots of the perturbed velocity vector fields, the perturbed velocity streamlines, and the interface perturbation. Real parts of fields and functions are shown. Each plot has its own range of values to better present the plot's features.

Figure 1

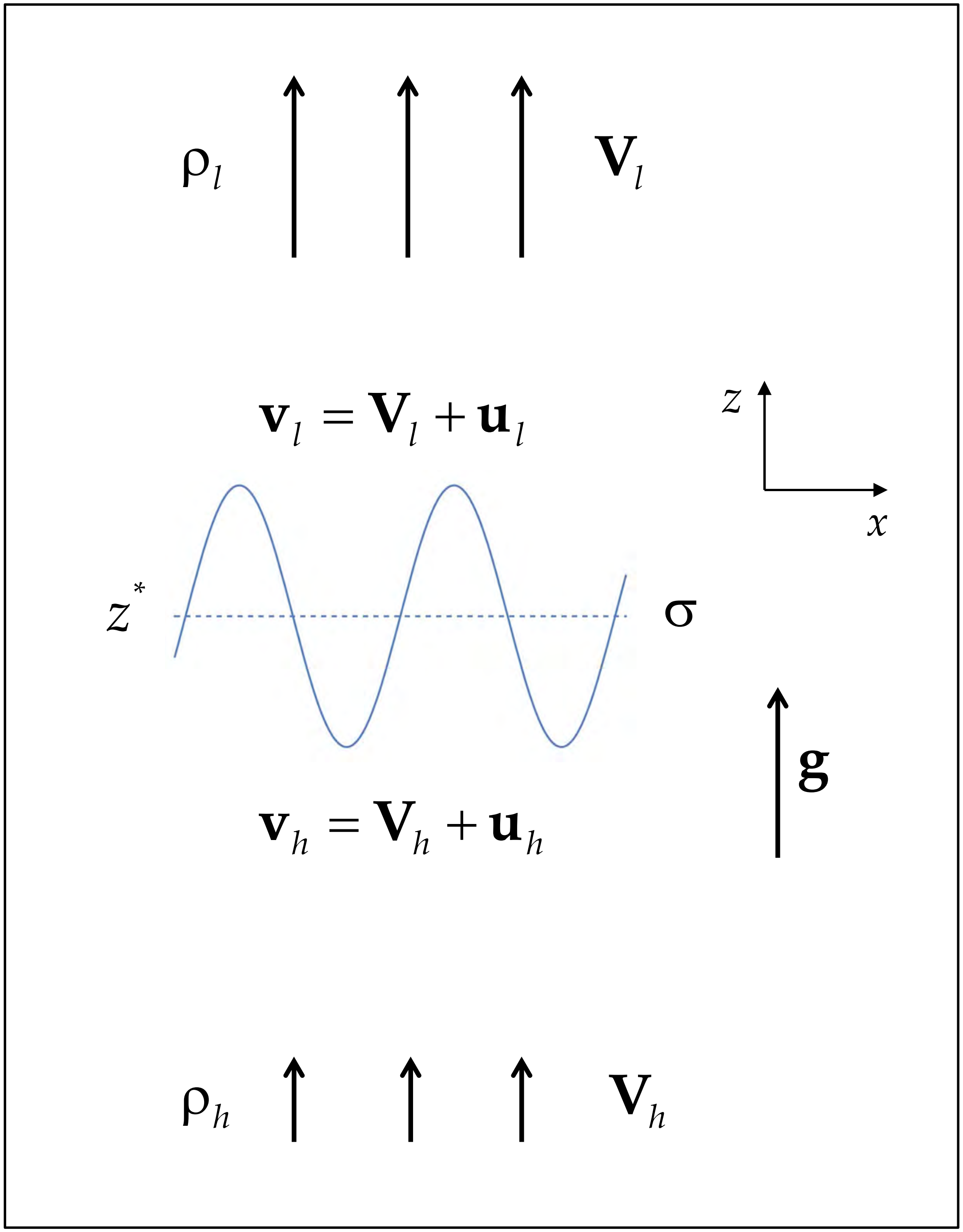

Figure 2

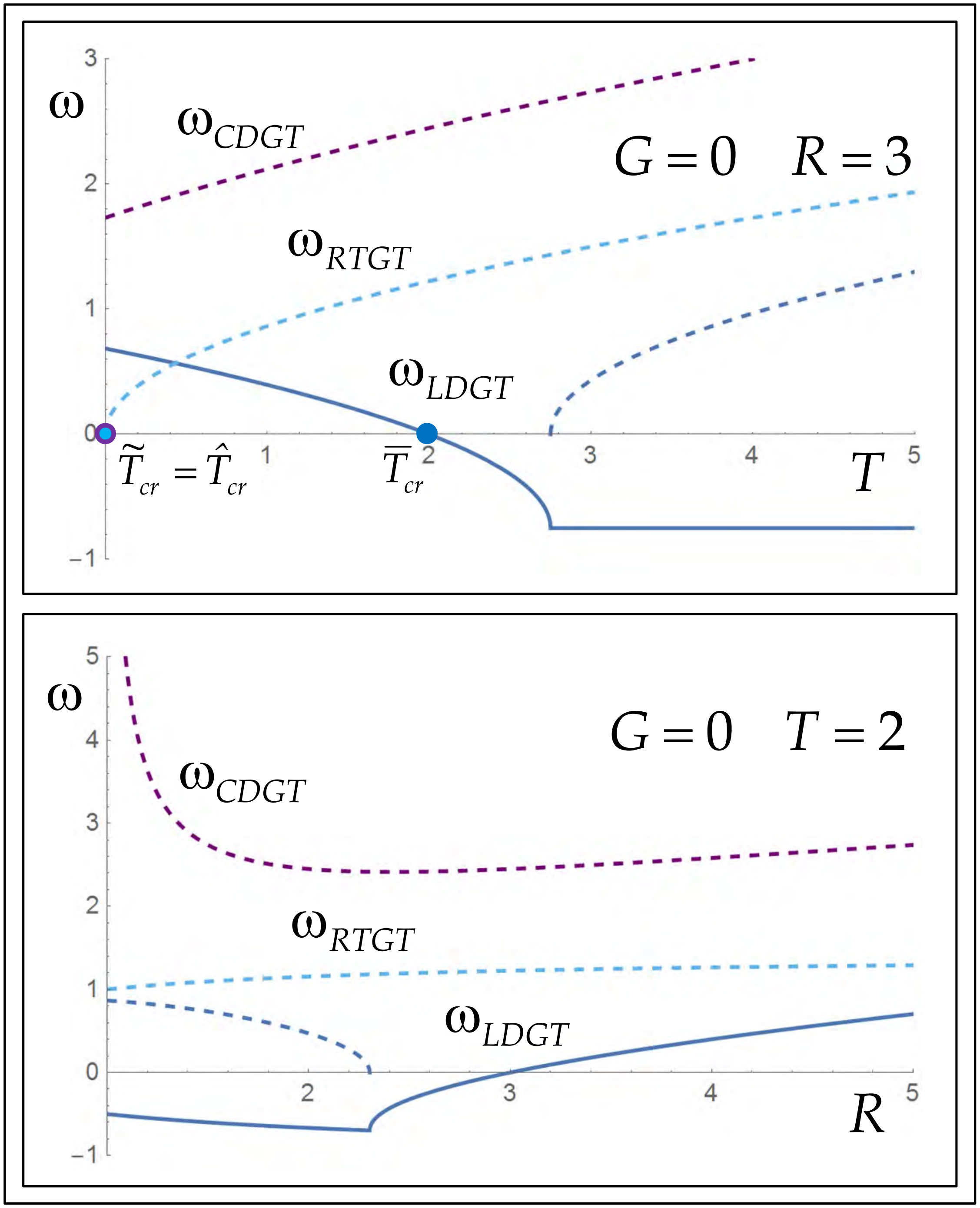

Figure 3a

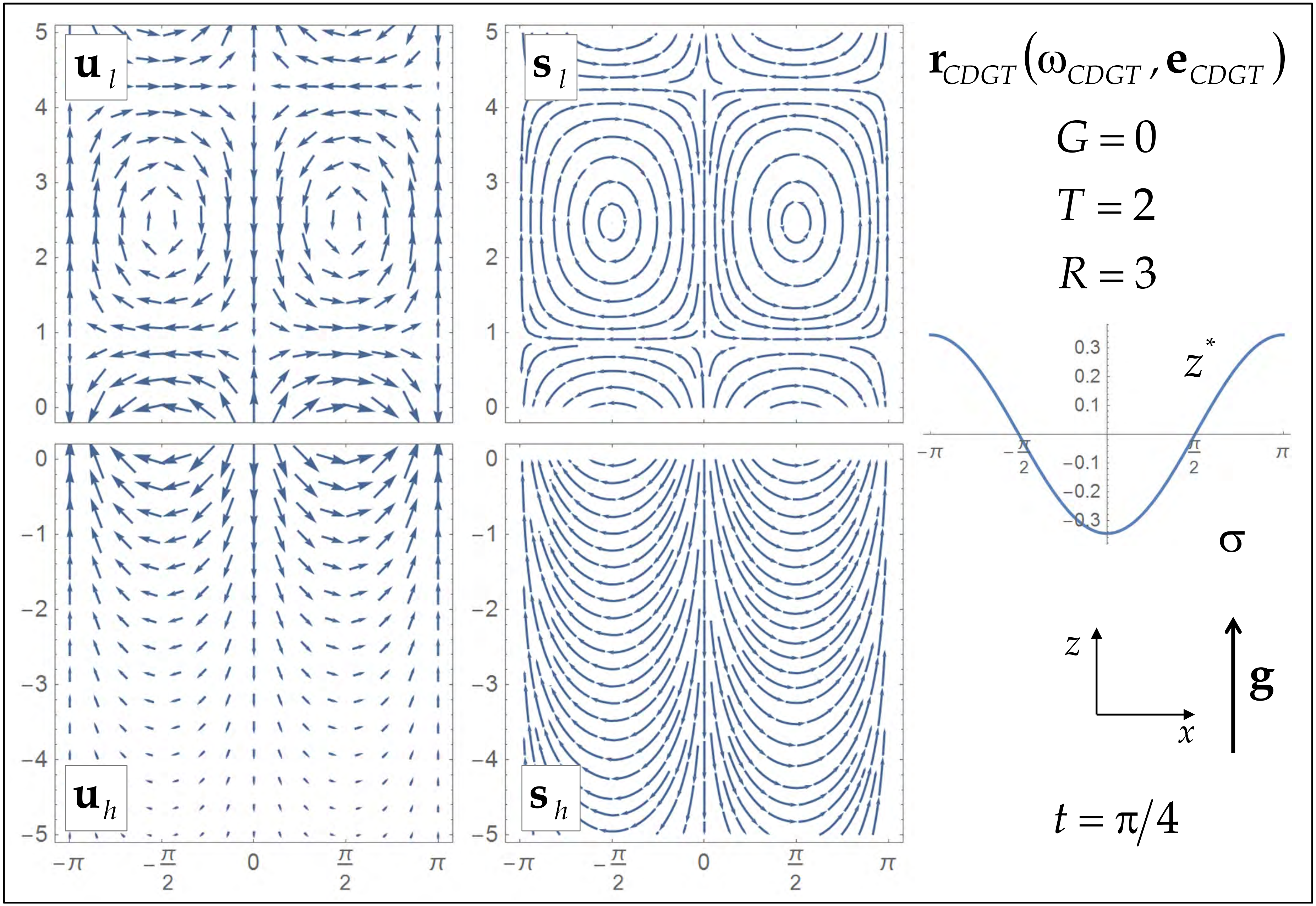

Figure 3b

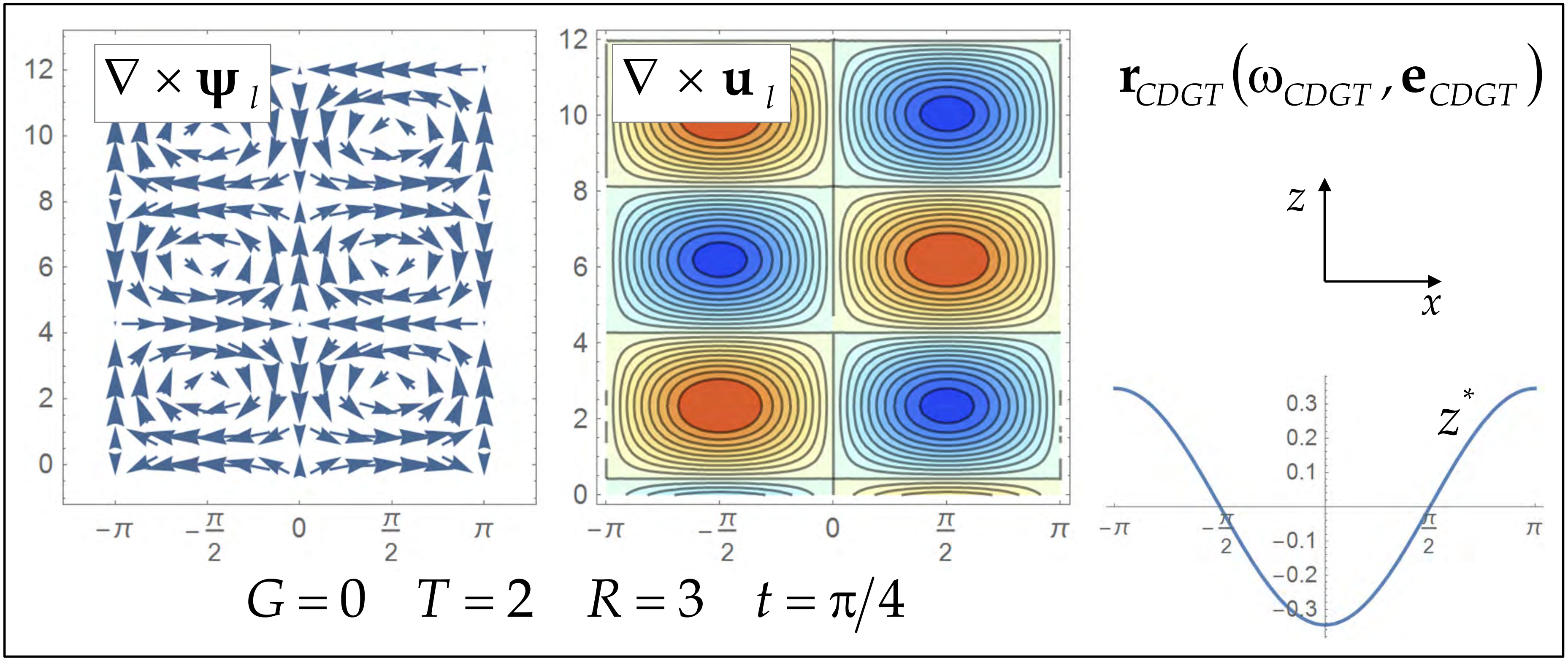

∇ × ψ l
∇ × u l
12
10
8
6
4
2
0
−π
−π/2
0
π/2
π
r CDGT (ω CDGT , e CDGT )
z
x
z*
0.3
0.2
0.1
−0.1
−0.2
−0.3
G = 0 T = 2 R = 3 t = π/4

Figure 4

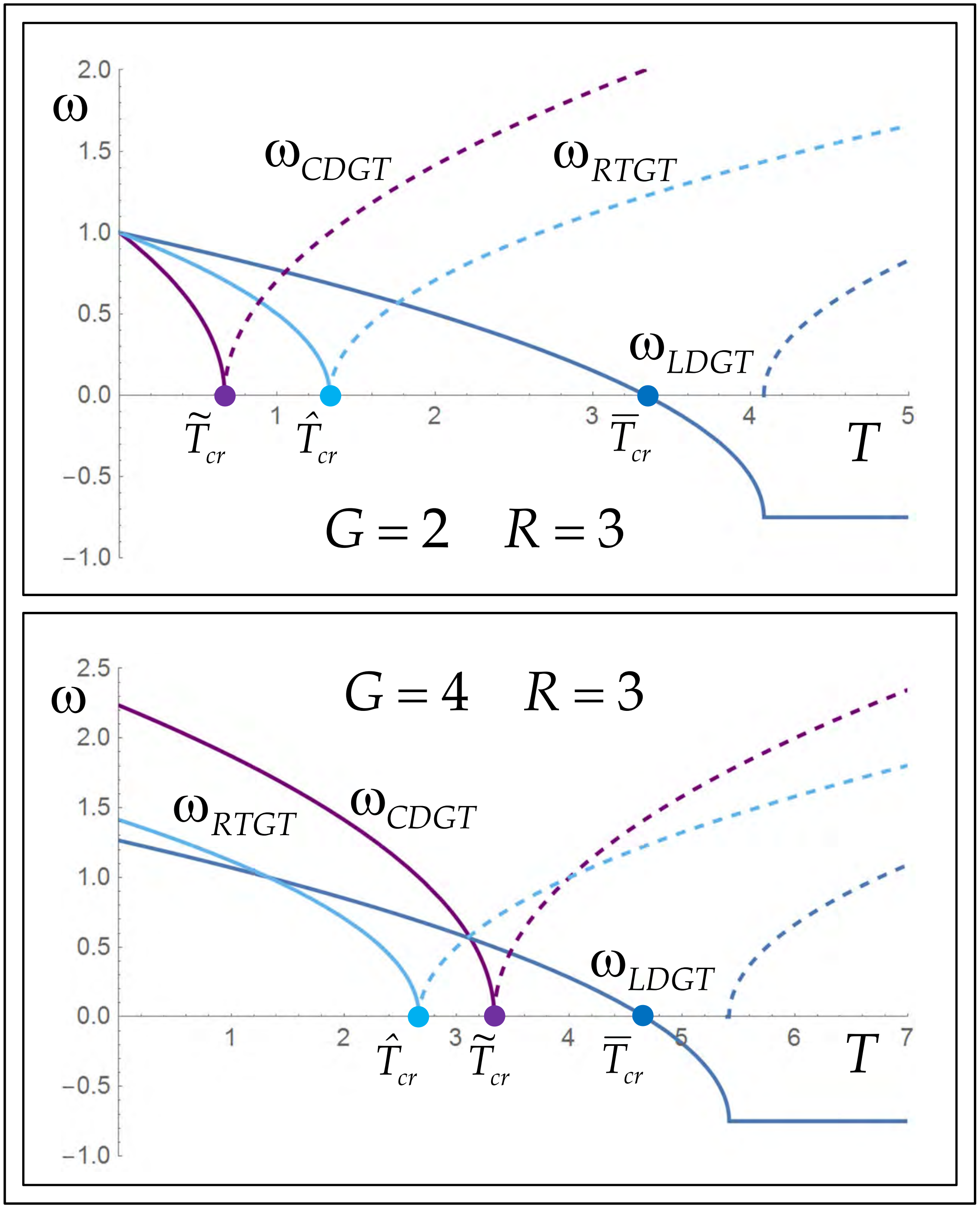

Figure 5

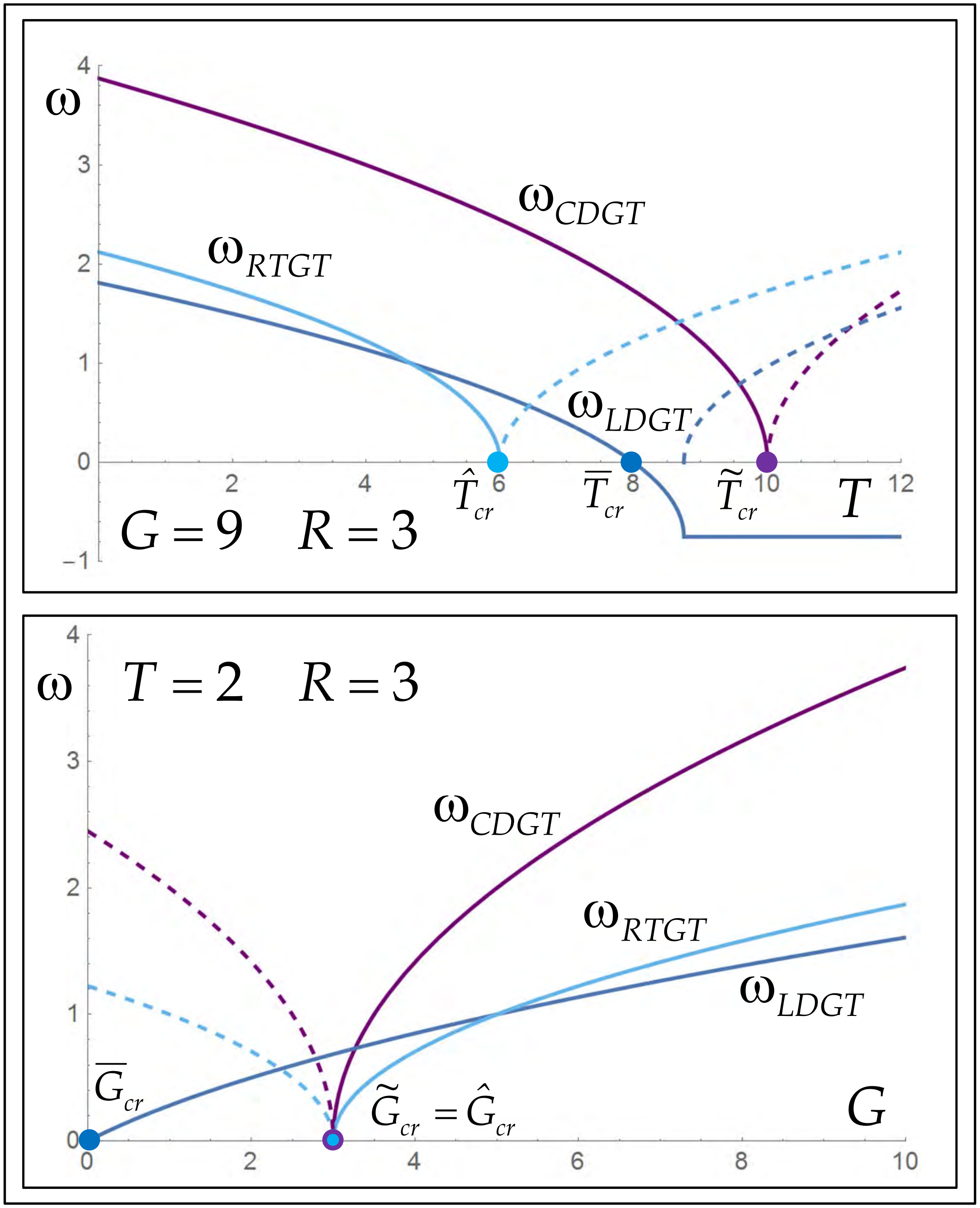

Figure 6a

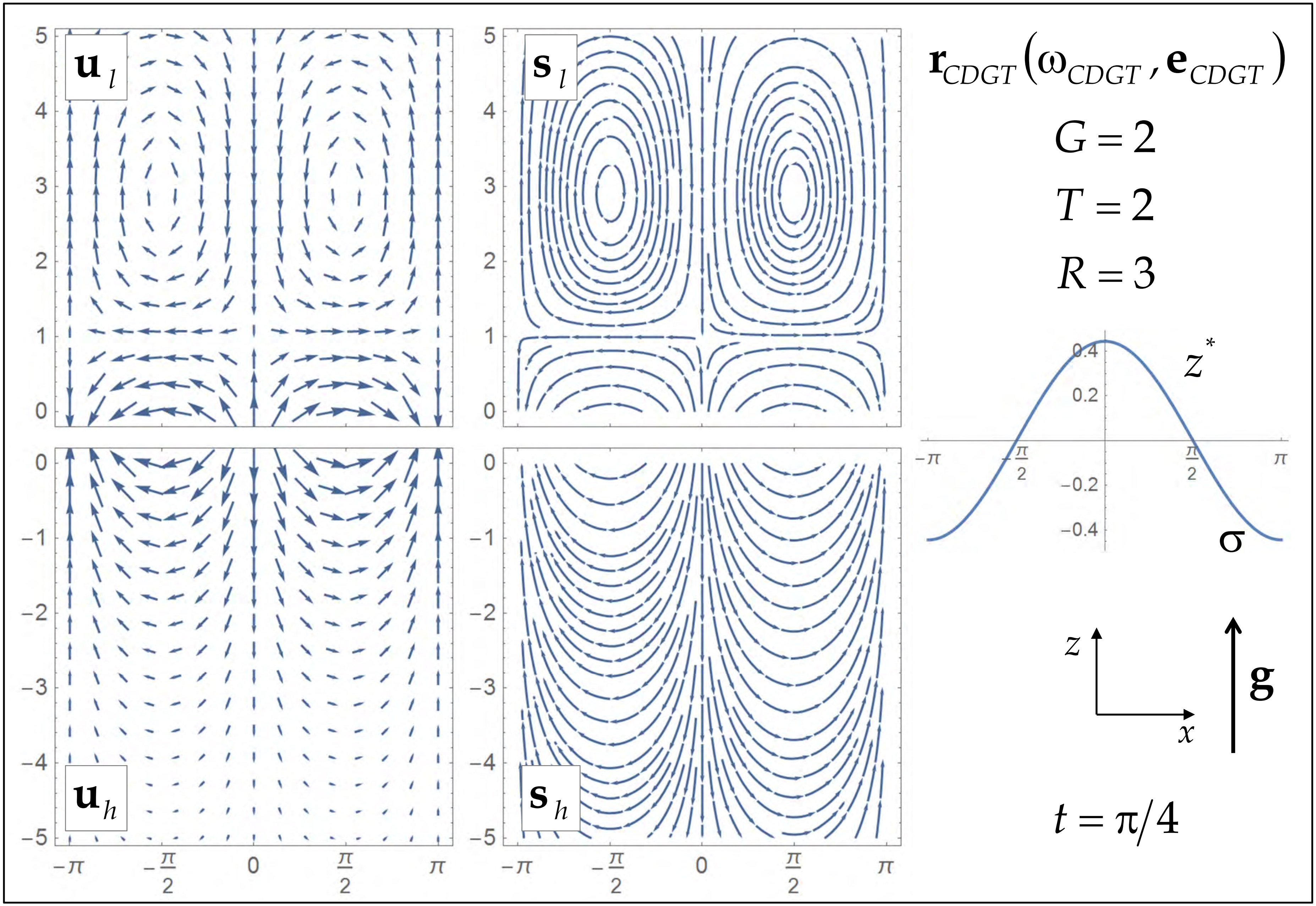

Figure 6b

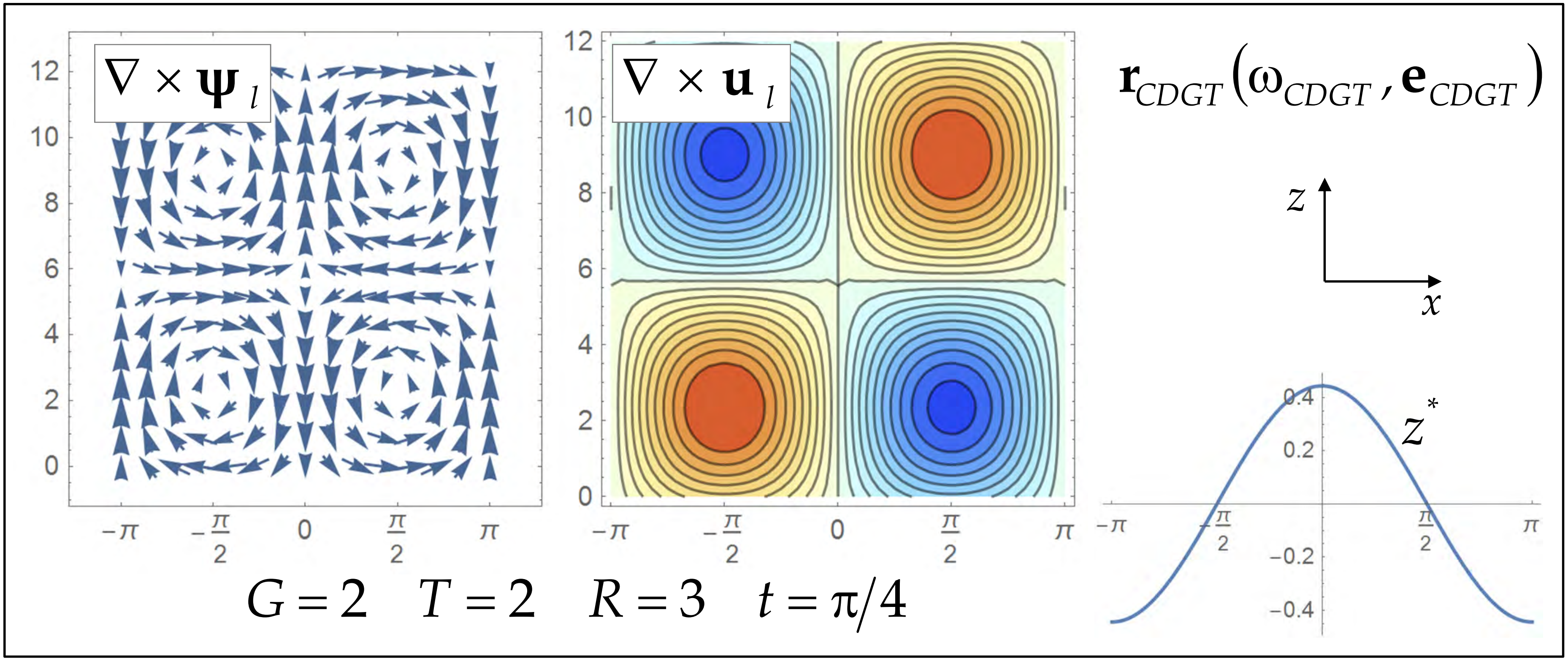

Figure 7a

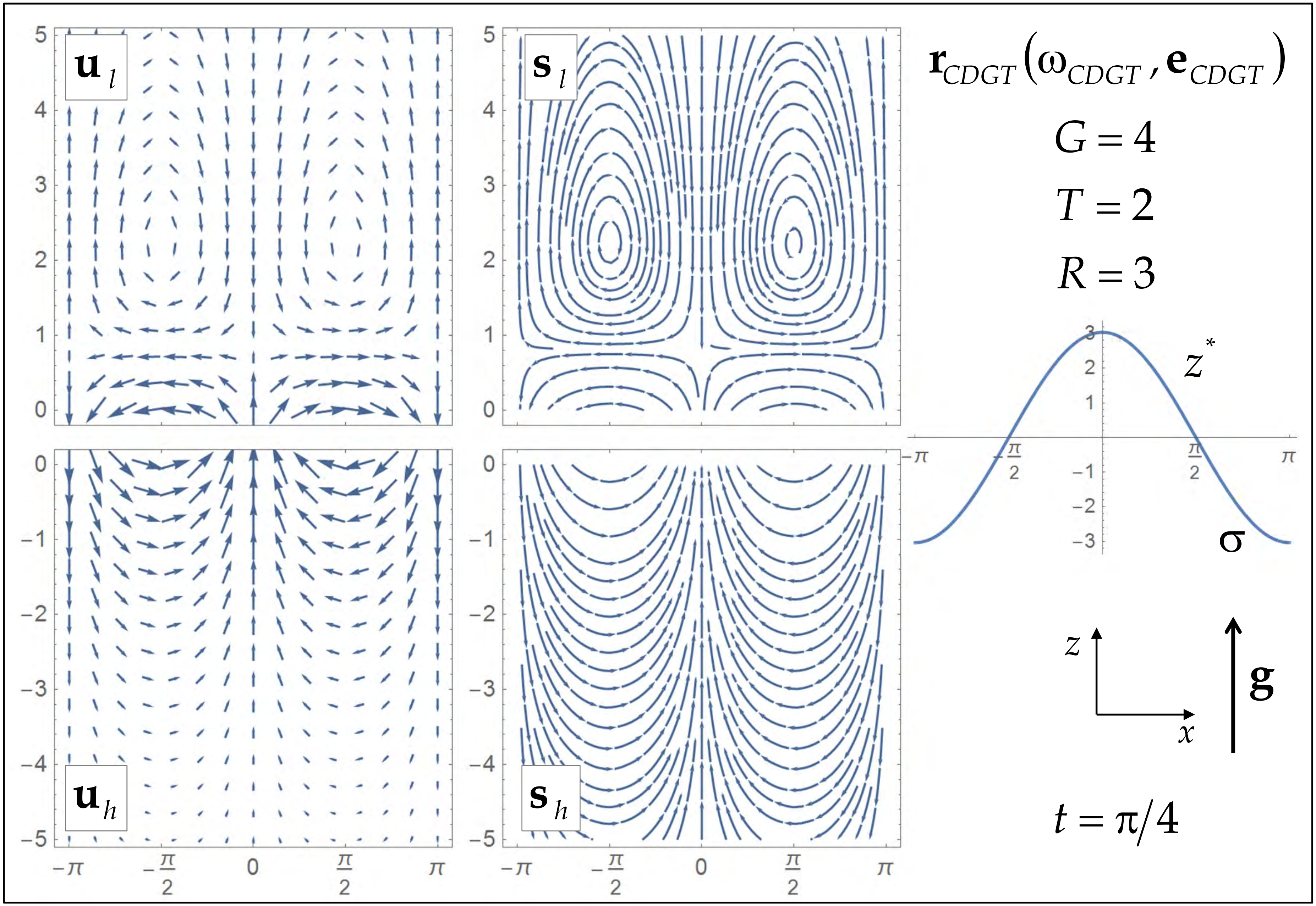

Figure 7b

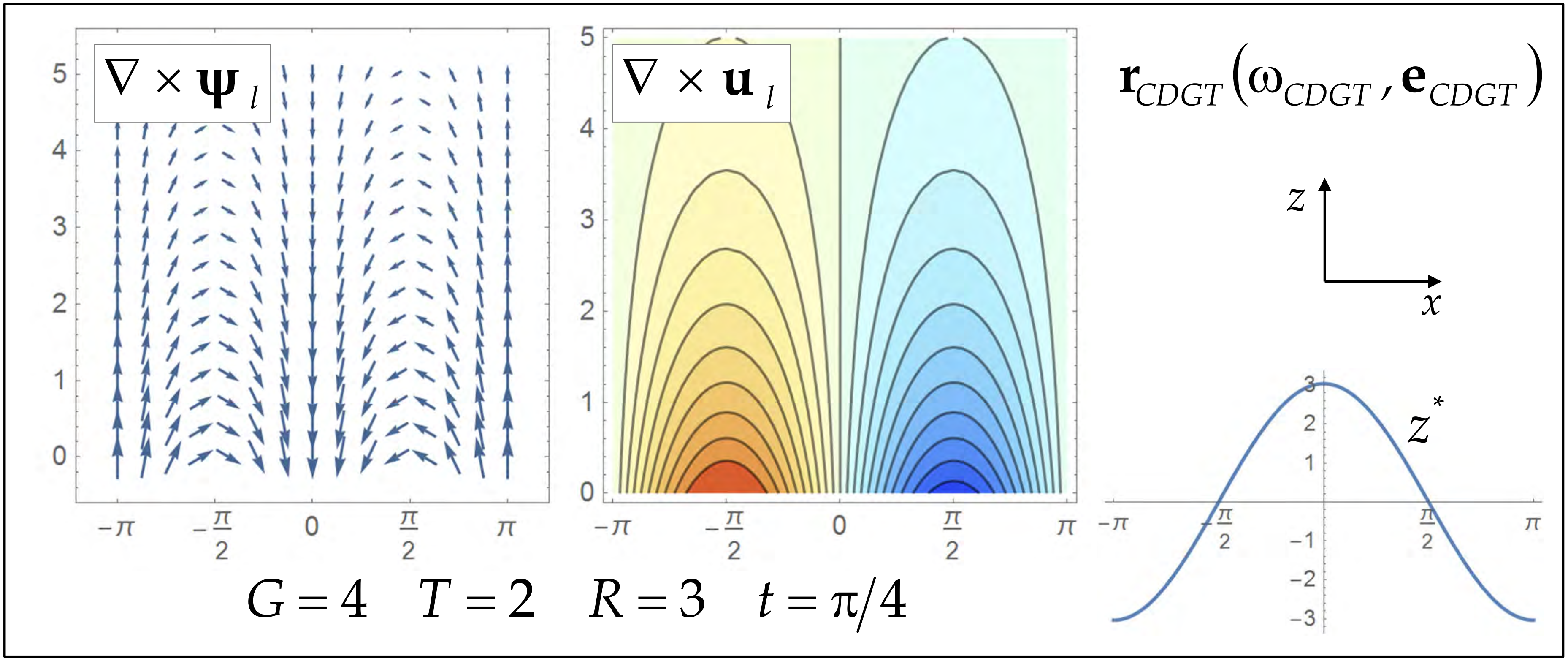

Figure 8a

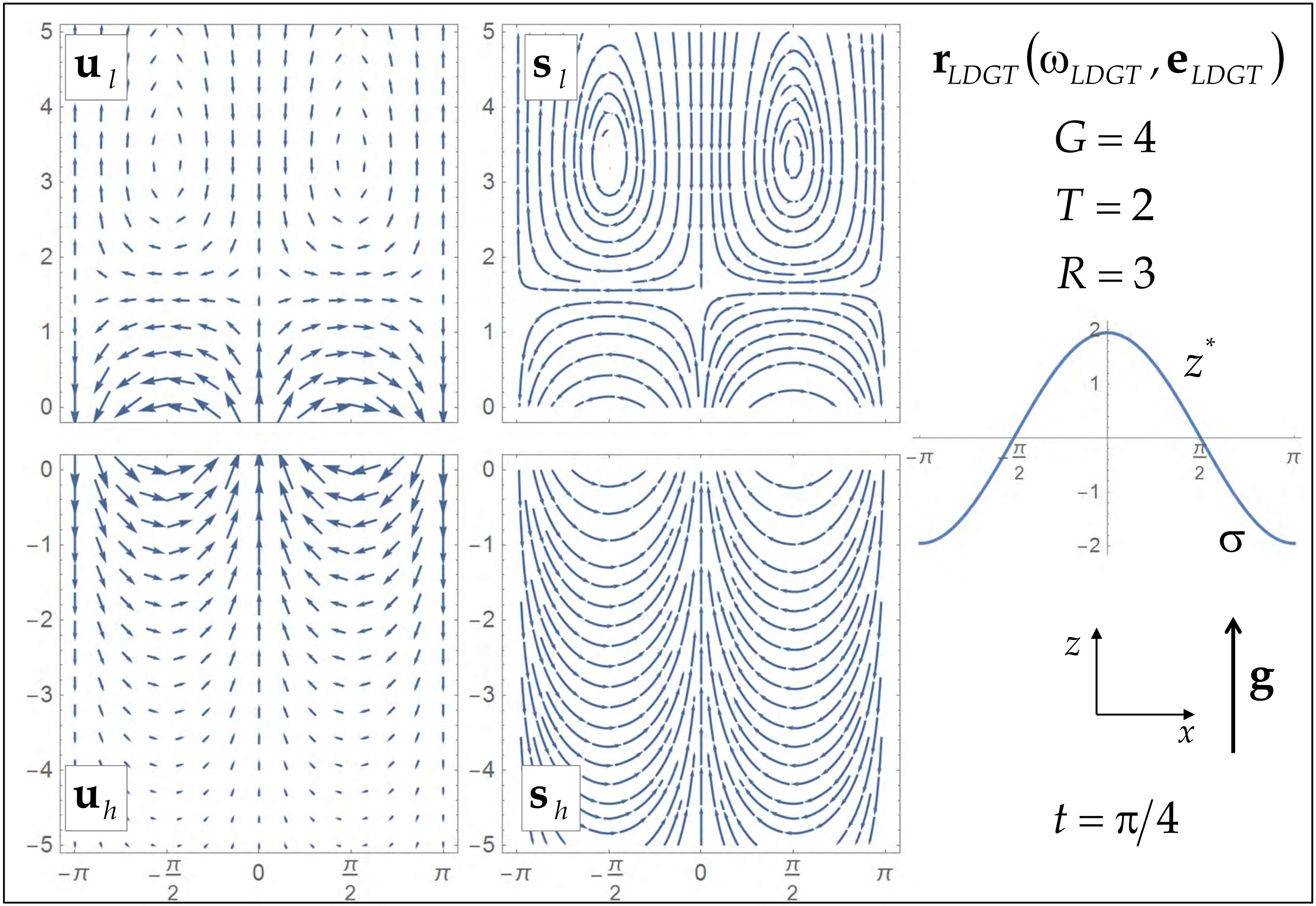

Figure 8b

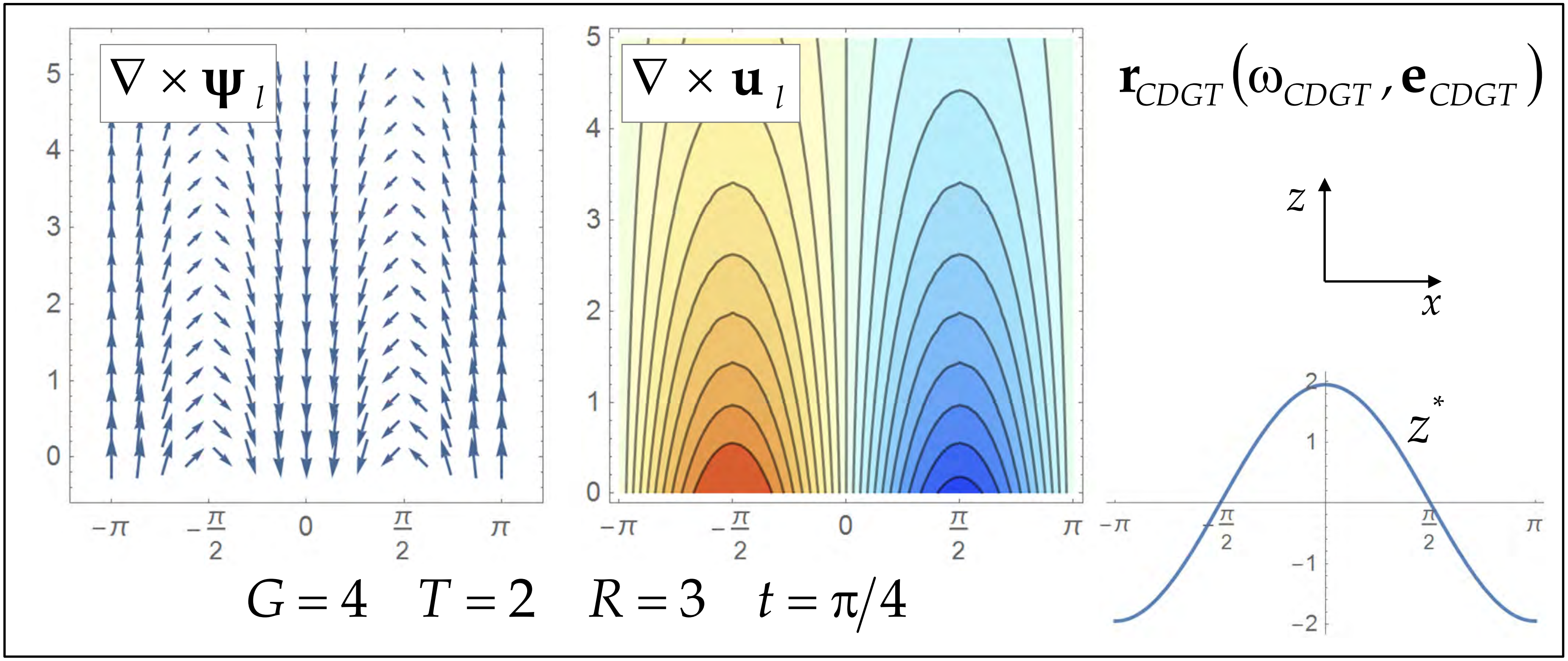

Figure 9

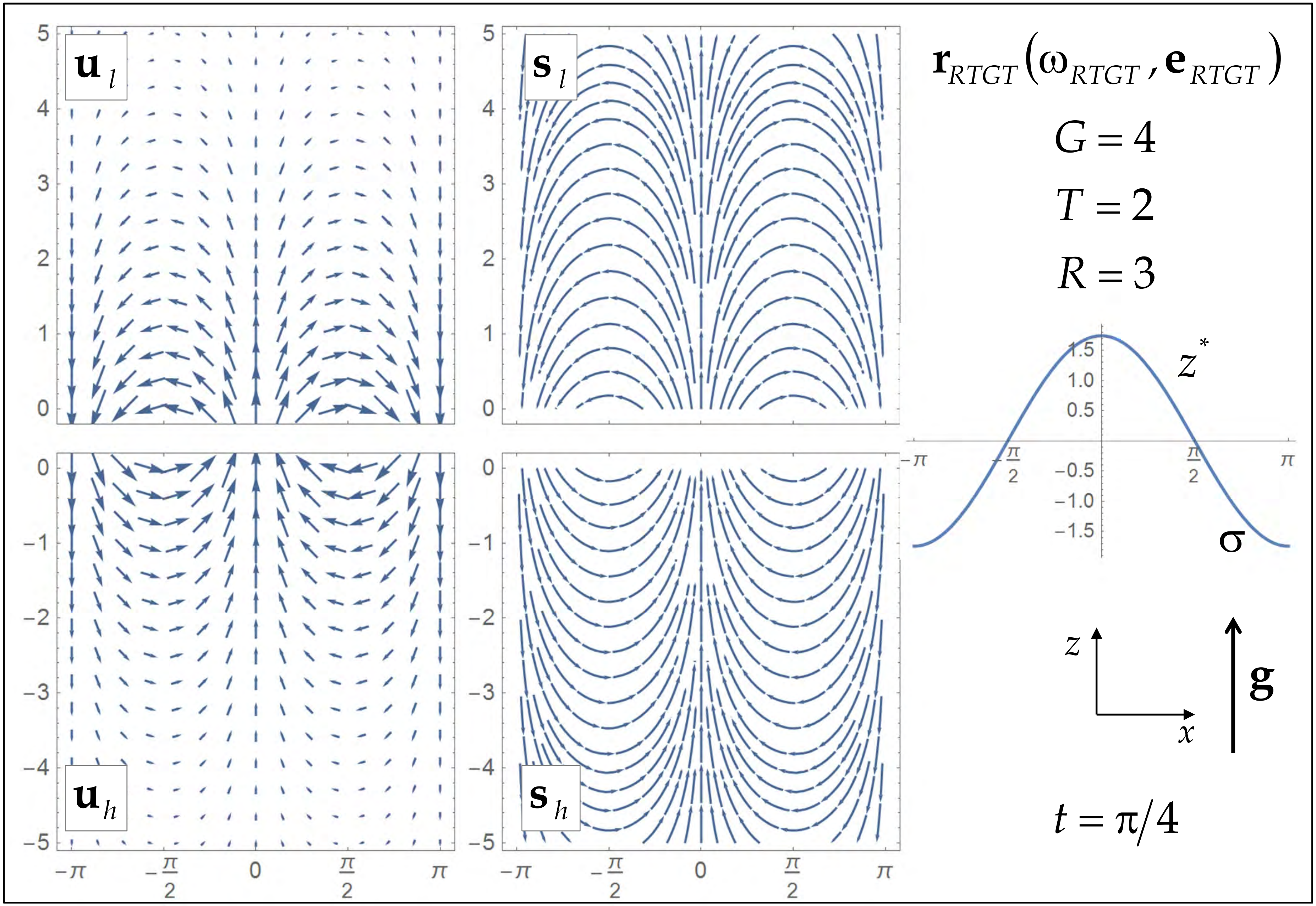